\documentclass[english,superscriptaddress,floatfix,twocolumn,oneside, amsmath,amssymb,amsfonts,aps,pra,floatfix]{revtex4-1}
\usepackage[english]{babel}
\usepackage[utf8x]{inputenc}
\usepackage{microtype}
\usepackage{graphicx}
\usepackage{siunitx}
\usepackage{xspace}
\usepackage[hidelinks]{hyperref}
\usepackage{bm}
\usepackage{bbold}
\usepackage{natbib}
\usepackage{csquotes}
\usepackage{amsmath}
\usepackage{qcircuit}

\newcommand{\cnot}{\textsc{cnot}\xspace}
\newcommand{\notgate}{\textsc{not}\xspace}
\newcommand{\swap}{\textsc{swap}\xspace}
\newcommand{\iswap}{$i$\textsc{swap}\xspace}
\newcommand{\dft}{\textsc{dft}\xspace}
\newcommand{\ps}{\textsc{ps}\xspace}
\newcommand{\eom}{\textsc{eom}\xspace}

\newcommand{\affA}{Department of Physics and Astronomy, Aarhus University, DK-8000 Aarhus C, Denmark}
\newcommand{\affB}{Aarhus Institute of Advanced Studies, Aarhus University, DK-8000 Aarhus C, Denmark}
\newcommand{\affC}{Hannover Center for Optical Technologies (HOT), Leibniz University Hannover, D-30167 Hannover, Germany.}

\begin{document}

\title{Reducing the amount of single-qubit rotations in VQE and related algorithms}

\author{S. E. Rasmussen}
\email{stig@phys.au.dk}
\affiliation{\affA}
\author{N. J. S. Loft}
\affiliation{\affA}
\author{T. Bækkegaard}
\affiliation{\affA}
\author{M. Kues}
\affiliation{\affC}
\author{N. T. Zinner}
\email{zinner@phys.au.dk}
\affiliation{\affA}
\affiliation{\affB}

\begin{abstract}
	With the advent of hybrid quantum classical algorithms using parameterized quantum circuits the question of how to optimize these algorithms and circuits emerges. In this paper we show that the number of single-qubit rotations in parameterized quantum circuits can be decreased without compromising the relative expressibility or entangling capability of the circuit. We also show that the performance of a variational quantum eigensolver is unaffected by a similar decrease in single-qubit rotations.
	We compare relative expressibility and entangling capability across different number of qubits in parameterized quantum circuits.
	High-dimensional qudits as a platform for hybrid quantum classical algorithms is a rarity in the literature. Therefore we consider quantum frequency comb photonics as a platform for such algorithms and show that we can obtain an relative expressibility and entangling capability comparable to the best regular parameterized quantum circuits.
\end{abstract}

\maketitle

\section{Introduction}

A large-scale, fault-tolerant, universal quantum computer is one of the end goals in the quest of quantum information technology. Despite the fact that we are closer to this significant milestone than ever before, it may take years before we actually conquer this. In fact current quantum technology supports only a couple of tens of qubits and a few hundred gate operations before the noise becomes too overwhelming. Nonetheless there is growing consensus that these Noisy Intermediate-Scale Quantum (NISQ) devices may find useful application in the near future \cite{Preskill2018}.

A strategy for optimizing the use of noisy quantum hardware is to divide the computational tasks between classical and quantum resources. Such a scheme is the development of hybrid quantum classical (HQC) algorithms. Examples of HQC algorithms are the quantum approximate optimization algorithm (QAOA) \cite{Fahri2014,Otterbach2017,Moll2018}, the quantum autoencoder (QAE) \cite{Romero2017}, the quantum variational error corrector (QVECTOR) \cite{Johnson2017}, classification via near term quantum neural networks (QNN) \cite{Fahri2018,Havlicek2019,Schuld2020}, quantum generative adversarial networks (qGAN) \cite{Dallaire-Demers2018,Lloyd2018,Zoufal2019,Zhu2019}, and the variational quantum eigensolver (VQE) \cite{Peruzzo2014,McClean2016,Omalley2016,Kandala2017,Cao2019,Barkoutos2018,McCaskey2019,Gard2020}. All these algorithms have in common that they share a quantum subroutine for producing parameterized trial states, where the parameters can be tuned to optimize a function value. Thus the performance of these algorithms depends on the configuration of the parameterized quantum circuit (PQC). This has led to several studies of circuit properties and capabilities \cite{Sim2019,Geller2018,Du2018,Benedetti2019,Hubregtsen2020}. In this paper, we characterize the power of a PQC by its entangling capability and so-called expressibility, introduced recently by Sukin Sim \emph{et al.} \cite{Sim2019}. 

In this article we focus on the effect of single-qubit rotations in PQCs. In particular we investigate whether it is possible to saturate the expressibility and entangling capability when gradually increasing the number of single-qubit rotations. While single-qubit rotations are often cheap to implement in the quantum circuit, they do sum up to many variational parameters for the classical optimization. If some of these variational parameters are redundant, the VQE result does not suffer by removing them from the classical optimization. In addition, the presence of redundant parameters slow down the optimizer and may get it stuck in a local minimum, preventing the VQE algorithm to converge on the correct value. With the current few-qubit VQE experiments, this has posed no problem, but for future applications to large systems beyond classical numerical methods, the amount of variational parameters in the classical part of the VQE algorithm may likely become the limiting factor. We therefore investigate how many rotations are necessary in order to run these algorithms without compromising the result. We do this by considering the expressibility and entangling capability \cite{Sim2019}, and investigate whether they saturate before we reach the maximum amount of single-qubit rotations. To verify that these metrics capture the capability of an actual HQC algorithm we also simulate a VQE for the same number of single-qubit rotations. In Section \ref{sec:Theory} we present the PQCs in relation to the VQE algorithm, and we introduce the two metrics we investigate. In Section \ref{sec:Results} we present our results of the metrics as well as the results of a VQE algorithm with four qubits run on the Lithiumhydride (LiH) molecule.

We also consider how the expressibility and entangling capability are affected when the number of qubits in the PQC is increased up to ten qubits, where previous investigations have only considered four qubits \cite{Sim2019}. In particular we consider how different entangling gates perform when the number of qubits is increased, in Section \ref{sec:NoQubits}.

A rather unexplored quantum computation scheme for HQC algorithms is systems of higher dimensions than two-level qubits. An example of such a scheme is quantum frequency comb (QFC) photonics. This scheme is a promising candidate for future quantum information technology, as the photons which constitute the qudits have a long lifetime \cite{Kues2017}. Another advantage is that it occurs naturally in high-bandwidth systems and can be controlled by commercially available telecommunications equipment, such as the electro-optic phase modulator (\eom) \cite{Wooten2000} or the Fourier-transform pulse shaper (\ps) \cite{Monmayrant2010,Weiner2011}. In Section \ref{sec:photonics} we consider high-dimensional qudits in QFC photonics as a platform for HQC algorithms.

In Section \ref{sec:conclusion} we present a summary and outlook for future work.

\section{Theory}\label{sec:Theory}

\begin{figure}
	\includegraphics[width=\columnwidth]{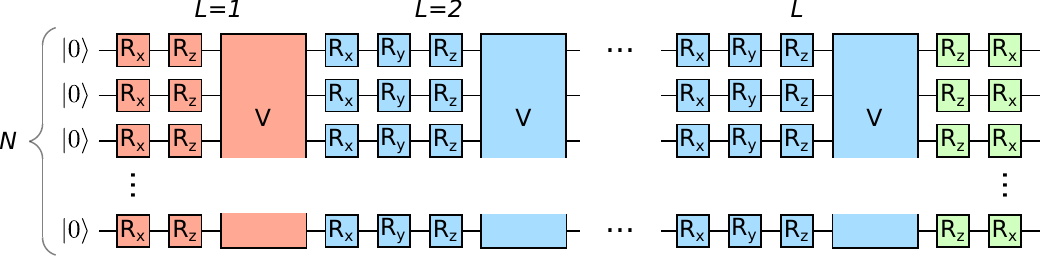}
	\caption{General implementation of a parameterized quantum circuit (PQC). Single-qubit rotations are denoted $R_x$, $R_y$, and $R_z$, while the multi-qubit block indicates potential multi-qubit operations. The first layer (red) consists of two single-qubit rotations on each qubit and a multi-qubit gate, $V$, the bulk layers (blue) each consists of three single-qubit rotations on each qubit and a multi-qubit gate. The circuit is completed with two single-qubit rotations on each qubit.}
	\label{fig:circuitTemplate}
\end{figure}

\begin{figure}
	\includegraphics[width=0.5\columnwidth]{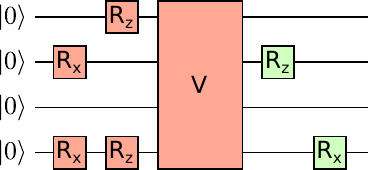}
	\caption{Example circuit with $N=4$ qubits, $L=1$ layer, and $m=6$ randomly configured single-qubit rotations.}
	\label{fig:circuit_example_6rots}
\end{figure}

In the VQE framework the quality of the results depends on how well a PQC, represented by a parameter-dependent unitary matrix, $\hat U(\bm \theta)$, can map an initial state of $N$ qubits, $|0\dots 0\rangle$, to the ground state of a Hamiltonian, $\hat H$. This is equivalent to optimizing $\bm \theta$ such that the expectation value of the Hamiltonian,
\begin{equation}
E(\bm\theta) = \langle 0 \dots 0|\hat U(\bm\theta)^\dagger \hat H \hat U (\bm\theta) |0 \dots 0\rangle,
\end{equation}
is minimized. The parameters, $\bm\theta$, of the circuit are usually related to single-qubit rotations. Consider the PQC in Figure \ref{fig:circuitTemplate}, which consists of $N$ qubits and $L$ layers. In the most general case, each layer performs an arbitrary rotation of each qubit on the Bloch sphere and an entangling operation, $V$.
In many quantum algorithms, one measures each qubit in the $z$-basis which by the same assumption makes final $z$-rotations irrelevant. However, in VQE the computation is finalized by measuring expectation values of Pauli operators, in which case all degrees of freedom matter for the qubit. Nevertheless, here we take the final rotations of the circuit (green) as a $z$- and $x$-rotation. Notice that the green gates are not considered a part of any layer and that the number of layers is equal to the number of $V$ operations. For now we assume that the entangling gates, $V$, have no variational parameters, meaning that the total number of parameters is equal to the number of single-qubit rotations,
\begin{equation}\label{eq:rotMax}
M = N(3L+1).
\end{equation}
For a VQE problem of fixed $V$, $L$, and $N$, we wish to investigate the quality of the PQC as a function of the number of single-qubit rotations, $m\in [0,M]$. However, as this will leave $2^M$ possible PQCs for each VQE problem, we select the PQC realizations randomly. For a given number of variational parameters, $m$, we pick a random circuit configuration with $m$ single-qubit rotations. This is equivalent to fixing $M-m$ rotation angles in Figure \ref{fig:circuitTemplate} to zero. An example circuit with $L=1$, $N=4$ and $m=6$ is shown in Figure \ref{fig:circuit_example_6rots}.
Producing many randomly chosen circuit configurations with $m$ rotations allow us to collect statistics about circuits with a certain `rotations filling degree'.

\subsection{Expressibility}

We wish to measure how well a given PQC performs in a HQC algorithm, such as VQE. One way to do this is to simulate a VQE algorithm for a specific system. However, we would like to characterize the PQC in a more general way, independent of the specific VQE problem. 
One way to do this is to calculate the expressibility (and entangling capability as presented in Section \ref{sec:entCap}) \cite{Sim2019}. The expressibility measures how close the PQC comes to uniformly mapping the initial state, $|0\dots 0\rangle $, to the entire Hilbert space.
This is done by comparing the probability distribution of the PQC with the probability of a uniform distribution in Hilbert space, i.e, an ensemble of Haar random state. Loosely, one may think of the expressibily as how many final states the PQC can reach.

It is important to note that the expressibility and entangling capability are not perfect metrics for the suitability of a PQC in a HQC \cite{Hubregtsen2020}. Some chemistry and condensed matter Hamiltonians contain specific symmetries, which can be solved accurately by a PQC respecting these symmetries, which is rated poorly by the expressibility and entangling capability metrics.

We calculate the expressibility following the approach in Ref. \cite{Sim2019}:
\begin{enumerate}
	\item Pick the gate operation, $V$, qubit number, $N$, and number of layers, $L$.
	\item For a given number of rotations, $m$, pick a random PQC  resulting in a parameterized unitary, $\hat U(\bm\theta)$.
	\item Calculate the expressibility of the circuit:
	\begin{enumerate}
		\item Uniformly sample $1000(N+1)$ sets of parameter vectors, $\bm \theta_1, \bm\theta_2$.
		\item Compute the overlap fidelity of the final states, $F=| \langle 0\dots 0 | \hat U^\dagger (\bm\theta_2) \hat U(\bm\theta_1) |0\dots 0\rangle |^2 $.
		\item Create a histogram over the fidelities, in order to estimate the probability distribution, $P(F)$, of the fidelities found in the previous step. Not that this estimation will depend on the number of bins in the histogram, $n_\text{bins}$. For the sake of consistency we set $n_\text{bins}=75$ as in Ref. \cite{Sim2019}.
		\item Compare $P(F)$ with the probability distribution achieved from an ensemble of Haar random states, $P_\text{Haar}(F) = (2^N -1) (1-F)^{2^N -2}$, using the same number of bins as in the previous step. We do this by computing the Kullback-Liebler divergence
		\begin{equation}
		\begin{aligned}
		\text{Expr} =& D_{KL}(P(F) || P_\text{Haar}(F))\\
		=& \sum_F P(F) \ln \left(\frac{P(F)}{P_\text{Haar}(F)}\right),
		\end{aligned}
		\end{equation}
		where the sum is taken over the bins. This measure is known as the expressibility of a PQC. 
	\end{enumerate}
	\item Calculate the relative expressibility 
	\begin{equation}\label{eq:relativeExpre}
	\mathcal{E} = -\ln \left[ \frac{\text{Expr}}{\text{Expr}(\text{Idle circuit})} \right] .
	\end{equation}
\end{enumerate}
Note that $\text{Expr} = 0$ for the most expressible circuit that reaches all final states uniformly, $P(F) = P_\text{Haar}(F)$. On the other hand, the least expressible circuit (the idle circuit with no gates) has $\text{Expr}(\text{Idle circuit}) = (2^N-1)\ln(n_\text{bins}) > 0$. This also illustrates the dependence on the chosen number of histogram bins. We therefore normalize our results with the expressibility of the idle circuit. To better resolve the most expressible circuits ($\text{Expr} \sim 0$), we take the logarithm and multiply by minus one in order to make the result positive. We call this the relative expressibility, see Equation \ref{eq:relativeExpre}.

We note that instead of calculating the expressibility via the Haar measure we could also have used other metrics, such as the qBAS score, where the probability distribution $P(F)$ is compared to the bars and stripes data set \cite{Benedetti2019b}.
Had we instead tried to maximize Expr, this could be considered training the circuit for approximating a given probability distribution. This unsupervised machine learning task is also known as generative modeling. Such a training model is an example of data-driven quantum circuit learning \cite{Benedetti2019b}.

\subsection{Entangling capability}\label{sec:entCap}

It is expected that HQC algorithms may have an advantage on similar classical algorithms in the future due to the fact that entanglement occurs naturally in the quantum mechanical part of the HQC algorithm.
Therefore we also wish to measure how well quantum circuits can produce entangled states. Reference \cite{Sim2019} proposes to use the entangling capability and defines it as the average Meyer-Wallach entanglement measure \cite{Meyer2002}
\begin{equation}
	Q(|\psi\rangle) = \frac{4}{n}\sum_{j=1}^{n} D(\iota_j (0) |\psi\rangle, \iota_j (1) |\psi\rangle),
\end{equation}
where the generalized distance, $D$, is
\begin{equation}
	D (|u\rangle, |v\rangle) = \frac{1}{2} \sum_{i,j} |u_iv_j - u_j v_i|^2,
\end{equation}
and the linear mapping $\iota_j(b)$ removes the $j$th qubit in the computational basis
\begin{equation}
	\iota_j(b) |b_1 \dots b_n\rangle = \delta_{bb_j} |b_1 \dots \hat b_j \dots b_n \rangle,
\end{equation}
where $B_j \in \{0,1\}$ and the 'hat' denotes the absence of the $j$th qubit. The average of $Q$ is then taken over a uniformly sampled set of parameters $\bm \theta$. We denote the entangling capability $\mathcal{C}$. The entangling capability lies between zero and one with zero being a non-entangling circuit and one representing a maximally entangling circuit. We wish to investigate whether the entangling capability saturates around the same amount of single-qubit rotations where the expressibility saturates.

\section{Results}\label{sec:Results}

The protocol we study is as follows. For each fixed number of rotations, $m$, we sample 100 random circuits and calculate their relative expressibility and entangling capability.
The PQC is considered saturated if both the relative expressibility and entangling capability converges after a given number of single-qubit rotations. The two metrics do not necessarily converge at the same number of single-qubit rotations, but, as we will see, this usually happens around the same point. One could argue that it this expected since we are only adding single-qubit gates, thus we are not adding any entangling gates to the circuit. Therefore a PQC which is saturated in the relative expressibility, should be able to apply the maximal entangling capability of the given $V$-gate. 

\begin{figure*}
	\includegraphics[width=\textwidth]{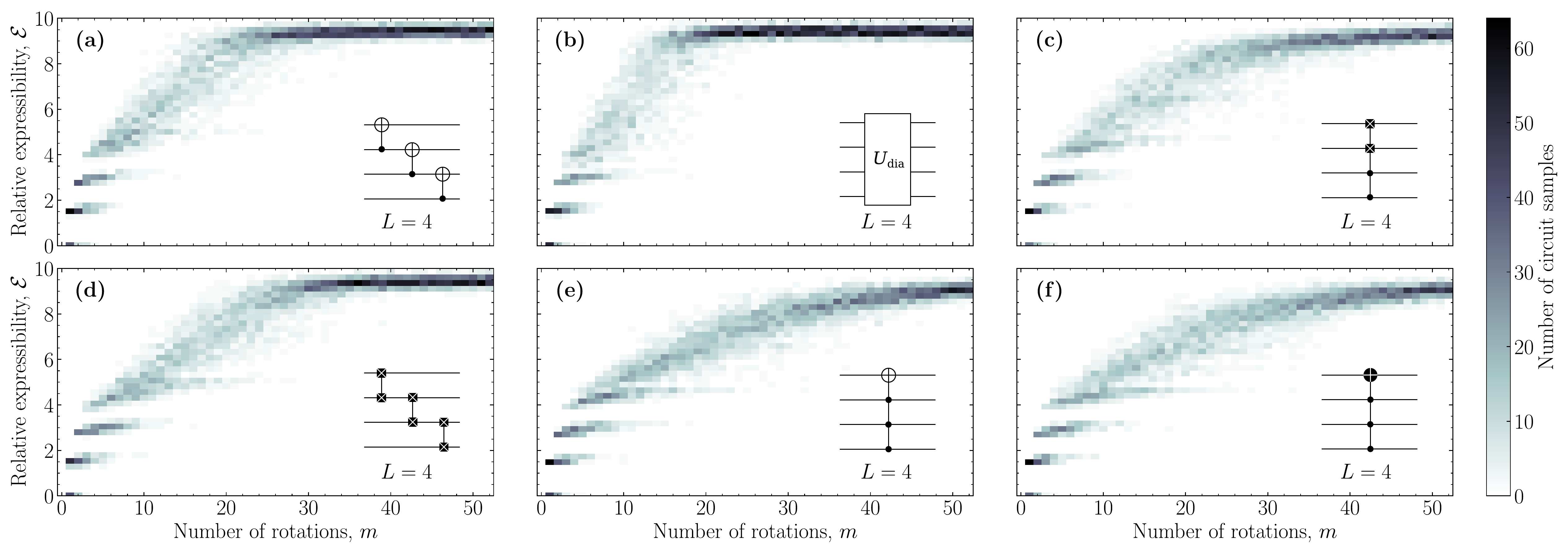}
	\caption{Relative expressibility of different circuits with $N=4$ qubits and $L=4$ layers as a function of the number of rotations. The inset in the right upper corner of each plot shows the entangling gate, $V$, used in each layer. \textbf{(a)} Three \cnot{}s, \textbf{(b)} diamond gate, \textbf{(c)} double controlled \iswap, \textbf{(d)} three \iswap{}s, \textbf{(e)} triple controlled \notgate, \textbf{(f)} triple controlled $i$\notgate.}
	\label{fig:exprL4}
\end{figure*}

\begin{figure*}
	\includegraphics[width=\textwidth]{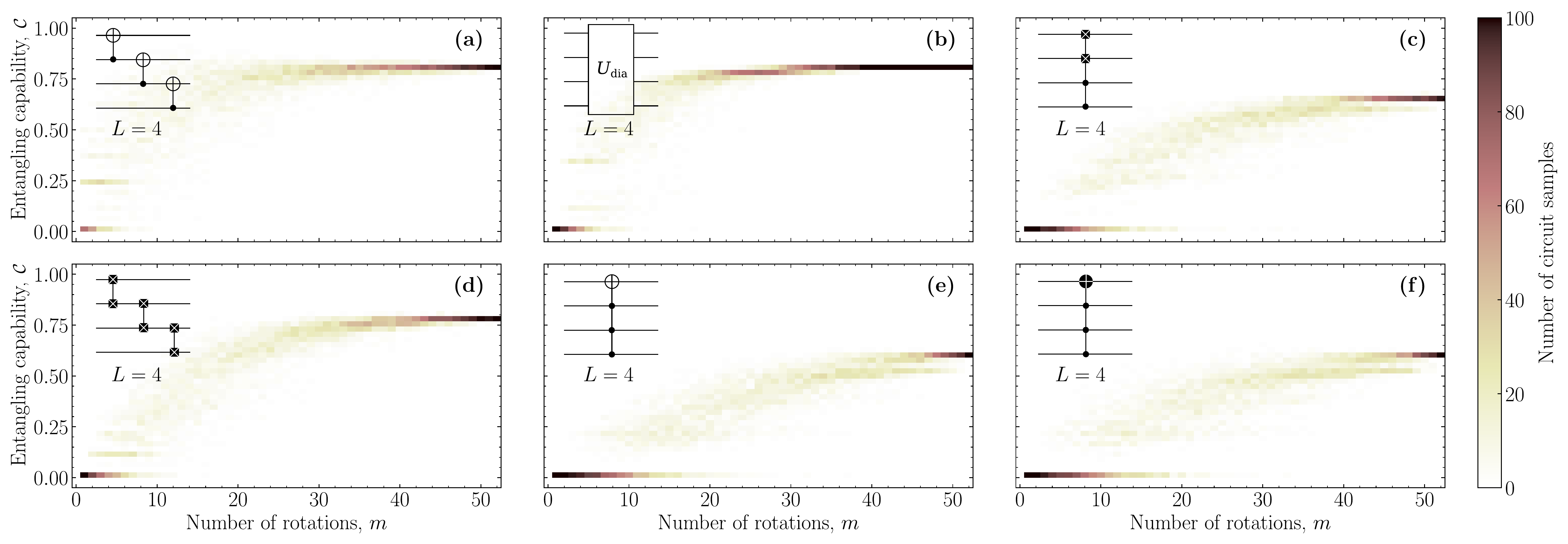}
	\caption{Entangling capability of different circuits with $N=4$ qubits and $L=4$ layers as a function of the number of rotations. The inset in the right upper corner of each plot shows the entangling gate, $V$,  used in each layer. \textbf{(a)} Three \cnot{}s, \textbf{(b)} diamond gate, \textbf{(c)} double controlled \iswap, \textbf{(d)} three \iswap{}s, \textbf{(e)} triple controlled \notgate, \textbf{(f)} triple controlled $i$\notgate.}
	\label{fig:entL4}
\end{figure*}

In Figure \ref{fig:exprL4} we have plotted a two-dimensional histogram of the relative expressibility as a function of the number of rotations, $m$, in the circuit. The calculations are done with $N=4$ qubits in $L=4$ layers, allowing for a maximum of $M=52$ single-qubit rotations according to Equation \ref{eq:rotMax}. The calculation is done for different entangling operations, $V$, displayed as the inset in each subplot. We present two-dimensional histograms of the entangling capabilities in Figure \ref{fig:entL4}, for the same cases. In order to assess the suitability of these metrics we also simulate a VQE algorithm with four qubits simulating the LiH molecule. The Hamiltonian of the LiH molecule is found in the STO-3G basis \cite{Pople1969} at the equilibrium of the molecule, which is then truncated to four qubits for the VQE calculation (see Section S2 in the Supporting Information for detail on the chemical construction of the molecular Hamiltonian). The results of the VQE calculation is compared to the results of a classical diagonalization of the reduced Hamiltonian.
These results are presented in Figure \ref{fig:vqeL4}. Section S3 in the Supporting Information contains results with $L=1,2,3$ and $N=6,8$.

Before discussing the details of each of the different $V$-gates we first consider some general features of the results. First of all, if no $V$-gates are present all product states can be represented using only three single-qubit rotations on each qubit. In the case of $N=4$ qubits this means that 12 single-qubit rotations are sufficient. This yields a relative expressibility of $\mathcal{E}=5.8$ and of course zero entangling capability. In fact all non-entangling circuits yields this relative expressibility and entangling capability.

Most of the plots show a 'stripe' pattern, which is especially pronounced for low relative expressibility or entangling capability. These stripes occur when the single-qubit rotations are placed on the same qubits. For $N=4$ we observe up to four stripes, the lowest stripe when all rotations are placed on one qubit, the second lowest when all rotations are placed on two qubits and so forth. For $N=6$ we observe up to six stripes, most easily seen on Figure S10(e) in the Supporting Information, one for each qubit. We also observe this stripe pattern in the VQE plots, see, e.g., Figure \ref{fig:vqeL4}. Here we interpret this as the PQC being able to only reproduce a restricted part of the Hilbert space of the molecule. These stripes are most dominant when the number of single-qubit rotations is low. Thus we conclude that when the number of single-qubit rotations is low, one must spread the single-qubit rotations evenly on all of the qubits, such that one qubit does not become over-saturated with redundant single-qubit rotations. When the number of single-qubit rotations increases the placement of these rotations is, however, less important.

Another common feature is that for eight qubits all circuits converge towards the same point. This is due to the choice of $n_\text{bins}$. When increasing the number of qubits the probability distribution of the Haar random states becomes more centered around zero. This means that for a fixed number of bins in the histogram, the first bin will dominate. This decreases the resolution of the expressibility. This could be fixed by increasing the number of bins when the number of qubits are increased. In the present paper we do not change the number of bins, as we want to compare systems with different numbers of qubits. Section \ref{sec:NoQubits} discusses different number of qubits in the PQCs. 

A final point is that a circuit that is converged in relative expressibility and entangling capability is not necessarily capable of finding any arbitrary state in the selected Hilbert space. The fact that we achieve convergence simply means that this particular choice of PQC has reached its limit. This can be seen by considering the minimum number of required parameters to specify an arbitrary $N$-dimensional state, which is $2(2^N)-2$ parameters, where the factor $2^N$ comes from the number of amplitudes of each basis state, the factor of two arises because the amplitudes are complex, and the subtraction of two parameters is due to global phase and normalization condition. Thus for the $N=4$ case the minimum parameter count is 30. This means that if a PQC saturates below $m=30$ it is not capable of reaching arbitrary states in the space.

\subsection*{$V=\textrm{\sc cnot}\textnormal{s}$}
\begin{figure*}
	\includegraphics[width=\textwidth]{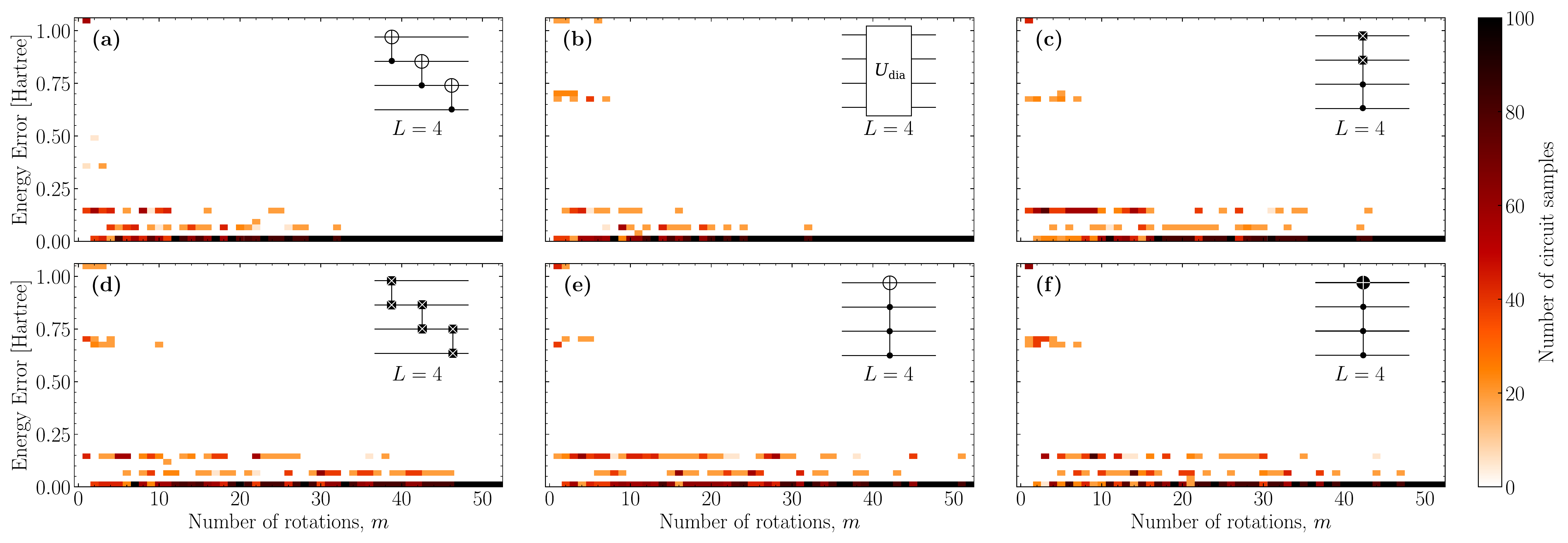}
	\caption{Energy error of the ground state LiH found using VQE of different circuits with $N=4$ qubits and $L=4$ layers as a function of the number of rotations. The energy found using VQE is compared to a FCI calculation, which yields the energy error. The inset in the right upper corner of each plot shows the entangling gate, $V$,  used in each layer. \textbf{(a)} Three \cnot{}s, \textbf{(b)} diamond gate, \textbf{(c)} double controlled \iswap, \textbf{(d)} three \iswap{}s, \textbf{(e)} triple controlled \notgate, \textbf{(f)} triple controlled $i$\notgate.}
	\label{fig:vqeL4}
\end{figure*}

One of the most frequently used entangling two-qubit gates is the \cnot gate. We therefore consider the case where $V$ is a \cnot between each consecutive pair of qubits. See Figure \ref{fig:exprL4}(a) for the case of four qubits and four layers. The relative expressibility converges towards $\mathcal{E}=10$ within 30 single-qubit rotations. For less layers the story is the same; the relative expressibility is converged at $m=30$ single-qubit rotations. This means that for one and two layers we do not saturate the relative expressibility, as we have less than 30 rotations in both cases. However, any more than three layers of \cnot{}s and 30 single-qubit rotations seems to be a waste of resources as the relative expressibility does not increase beyond this. This should be related to the fact that the minimum parameter count is 30 in this case, however, this does not mean that this PQC configuration is the most optimal as we will see when we discuss the diamond gate.
From Figure \ref{fig:entL4}(a) we observe that the entangling capability saturates at around 0.8 at approximately 30 single-qubit rotations.
The simulation of a VQE  algorithm solving the Hamiltonian of LiH for the ground state energy can be seen in Figure \ref{fig:vqeL4}(a). Here we observe that the energy error (compared to a classical eigenvalue calculation) saturates at less than 0.05 Hartree for around 30 single-qubit rotations.

For $N=6$ qubits (see Supporting Information) we observe the same tendencies, however, the relative expressibility converges towards $\mathcal{E}=9$. This happens at around 40 single-qubit rotations, which can be obtained for just two layers. This is significantly smaller than the minimum parameter count which is 126 for six qubits. For the $N=6$ results the VQE algorithm solves the ground state of BeH$_2$.

Turning towards the case of eight qubits we again observe a saturation of the relative expressibility and entangling capability (see Supporting Information). Convergence is reached after approximately 30 single-qubit rotations at a relative expressibility $\mathcal{E} =7.5$, lower than for both four and six qubits, and the same as for no entangling $V$ gate. This fact is due to the number of bins used in the definition of the relative expressibility, see Section \ref{sec:NoQubits} for further discussion.
The VQE calculations are, for the case of $N=8$, done for the ground state of OH, and it is clear to see a convergence of the energy error before reaching the maximum number of rotations. This is an important point of our work.

\subsection*{$V=i\textrm{\sc swap}\textnormal{s}$}
An entangling version of the \swap gate is the \iswap gate. The \iswap gate has an entangling power equivalent to the \cnot gate, and occurs naturally in systems with $XY$-interaction or Heisenberg models, such as solid state systems \cite{Tanamoto2008,Tanamoto2009}, superconducting circuits \cite{Zagoskin2006}, and in cavity mediated interaction between spin qubits and superconducting qubits \cite{Imamoglu1999,Benito2019,Blais2004}. In Figure \ref{fig:exprL4}(d) we present the relative expressibility where the $V$ gate is an \iswap gate between each qubit. The relative expressibility of this configuration goes towards the same as for the \cnot, however, it converges a bit slower. For four qubits the difference is about five rotations. However, for six qubits it is more pronounced with a difference of around 20 single-qubit rotations. The behavior for eight qubits and the \iswap gates is the same as for the \cnot gate.

The entangling capability in \ref{fig:entL4}(d) saturates a bit below 0.8 and the saturation point is later than for the \cnot gates as for the relative expressibility. However, in the VQE simulation we observe an even slower convergence, which is not reached before 45 rotations. All in all the \iswap gate performs worse than the \cnot gate.

\subsection*{$V=\textrm{\sc diamond}$}

An example of an entangling four-qubit gate is the diamond gate \cite{Loft2019}, which is an entangling swapping gate with two controls, which must be in an entangled state in order to control the swapping operation. The diamond gate is difficult to synthesize into one- and two-qubit gates. A decomposition into standard gate requires 21 gates, while it can be done with only \cnot gates and single-qubit gates using 42 and 49 gates, respectively. It does, however, naturally occur in superconducting circuit schemes for quantum information processing. See Section S1 in the Supporting Information for more details.

In Figure \ref{fig:exprL4}(b) we present the relative expressibility for circuits with four qubits, four layers and $V=U_\text{dia}$, where $U_\text{dia}$ is the diamond gate. We see that the relative expressibility converges rapidly towards $\mathcal{E}=10$, which it reaches within 20 single-qubit rotations, which is less than for both the \cnot and \iswap gates in subfigures (a) and (d) and less than the minimum number of parameters. We can therefore conclude that even though the PQC has saturated, it is incapable of reaching any arbitrary state in Hilbert space. This also means that other circuits that saturated at the same relative expressibility must be incapable as well.
In the case of the entangling capability (Figure \ref{fig:entL4}(b)) and VQE simulation (Figure \ref{fig:vqeL4}(b)) we also find the saturation at around 20 single qubit rotations.

We observe the same behavior for six qubits; when the diamond gate is used as an entangling gate, convergence is reached faster than for the \cnot and \iswap gates. For eight qubits the performance of the diamond gate is reduced to the same level as the two other gates.

We also see that the stripe patterns are less pronounced in the case of the diamond gate. This is because the gate is highly entangling, mixing the qubits and thus blurring out the stripes.

\subsection*{$V=\textrm{\sc Multi-qubit gates}$}

In order to investigate whether the diamond gate performs well simply because it is a multi-qubit gate, we consider a few other multi-qubit gates. A well known three-qubit gate is the Toffoli gate, a \notgate gate with two control qubits. Together with the Hadamard gate it constitutes a universal set of quantum gates \cite{Toffoli1980,Nielsen2010}. The Toffoli gate can be expanded using an $n$-controlled \notgate gate \cite{Isenhower2011,Molmer2011,Shi2018,Wang2001}. Another version of the $n$-controlled \notgate gate is the $n$-controlled $i$\notgate gate, which is a controlled \notgate gate which acquires a phase of $i$ if the \notgate gate is activated. The $n$-controlled \notgate gate can readily be implemented in systems where this phase occurs naturally, such as superconducting circuits \cite{Rasmussen2020a}. Related to the Toffoli gate is the Fredkin gate \cite{Fredkin1982}, a controlled \swap gate. The Fredkin gate can also be expanded using a \swap gate with multiple controls. Due to the fact that the Fredkin gate itself is difficult to implement we consider instead a multiple controlled \iswap gate, as it occurs naturally in many solid state system used to implemented quantum information schemes \cite{Rasmussen2020b,Pedersen2019,Baekkegaard2019}.

We consider the $n$-bit \iswap gate, the $n$-bit Toffoli, and the $n$-bit $i$Toffoli as the $V$ gate one at a time. The result using these gates can be seen in subfigures (c), (e), and (f) of figure \ref{fig:exprL4}, respectively, and in the Supporting Information for six and eight qubits. Common for all three gates are that they converge towards the same relative expressibility as the other entangling gates, namely $\mathcal{E}=10$ for four qubits. However, they are much more dependent on the number of rotations, as the convergence point is only reached for the maximum amount of rotations. The main difference of the three gates is that the double controlled \iswap in Figure \ref{fig:exprL4}(c) converges a bit faster than the multiple controlled \notgate gates. This is however, not something which can be seen for six qubits, where all three multi-qubit gates converge at the same rate. Thus the faster convergence of the double controlled \iswap is probably due to the fact that it has twice as many target qubits compared to the multiple controlled \notgate gates, something which is less significant for more qubits. An interesting thing to notice is the fact that the diamond gate seems to converge much faster than the double controlled \iswap gate (compare subfigure (b) to (c)). This is despite the fact that both gates are swapping gates with two controls. This is probably because the controlling part of the diamond gate requires a superposition state in order to activate the swap, compared to the multiple controlled \iswap gate which does not require a superposition.

In terms of entangling capability, multiple control gates perform only at an average level with a maximal entangling capability of 0.5, see Figure \ref{fig:entL4}(c), (e), and (f). As for the relative expressibility we need the maximum number of rotations for the entangling capabilities to reach its maximum.
The VQE simulation in Figure \ref{fig:vqeL4}(c), (e), and (f) shows the same tendency, i.e., that the saturation point is reached close to the maximum number of single-qubit rotations.

For six qubits we also observe that the three multi-qubit gates just barely outperform the non-entangling gates. This could be due to the fact that most of the qubits become control qubits when the number of qubits increases. This means that the multi-qubit gates will act trivially on more states as the number of qubits increases.

We conclude that the entangling capability and relative expressibility saturates for approximately the same number of qubit rotations as the VQE simulation saturates. Due to the computational complexity, we only calculate two-dimensional entangling capability histograms for up to  six qubits in the Supplementary Information.
However, we do expect the same results in the eight qubit case as in the four and six qubits cases. In the following section we do, nonetheless, consider the entangling capability and the relative expressibility as the number of qubits increases, but only for the case of a fully saturated PQC.

Table \ref{tab:overview} presents a comparison of the different $V$-gates used in the analysis. The comparison is made after the saturation point is reached for the diamond gate with two layers and 25 single-qubit rotations for both four and six qubits. We do not compare the eight qubit case as the expressibilities are the same for all $V$-gates. See Section \ref{sec:NoQubits} for a discussion of why this is the case.
We note that there may be other entangling gates which outperform the gates we have considered here.

\begin{table}
	\caption{Comparison of relative expressibility and entangling capability for the different multi-qubit $V$-gates. The comparison is done after the saturation point of the diamond gate is reached with $L=2$ layers and $m=25$ rotations. Expressibility and entangling capabilities are taken as averages over the sample size.
	We compare both four and six qubit, but not for eight qubits since the expressibilities are the same for all $V$-gates due to our choice of bins.}
	\label{tab:overview}
	\begin{tabular}{lcccc}
		\hline
		 & \multicolumn{2}{c}{$N=4$} & \multicolumn{2}{c}{$N=6$} \\
		$V$-gate & $\mathcal{E}$ & $\mathcal{C}$ & $\mathcal{E}$ & $\mathcal{C}$ \\
		\hline
		None & 5.8 & 0 & 6.6 & 0 \\
		\cnot{} & 8.7 & 0.68 & 7.9 & 0.70 \\
		\iswap{} & 8.3 & 0.61 & 7.2 & 0.42 \\
		\textsc{diamond} & 9.4 & 0.79 & 8.4 & 0.76 \\
		\textsc{Multi-qubit} & 7.3 & 0.35 & 6.7 & 0.05 \\
		\hline
	\end{tabular}
\end{table}

\section{Increasing the number of qubits}\label{sec:NoQubits}

\begin{figure*}
	\includegraphics[width=\textwidth]{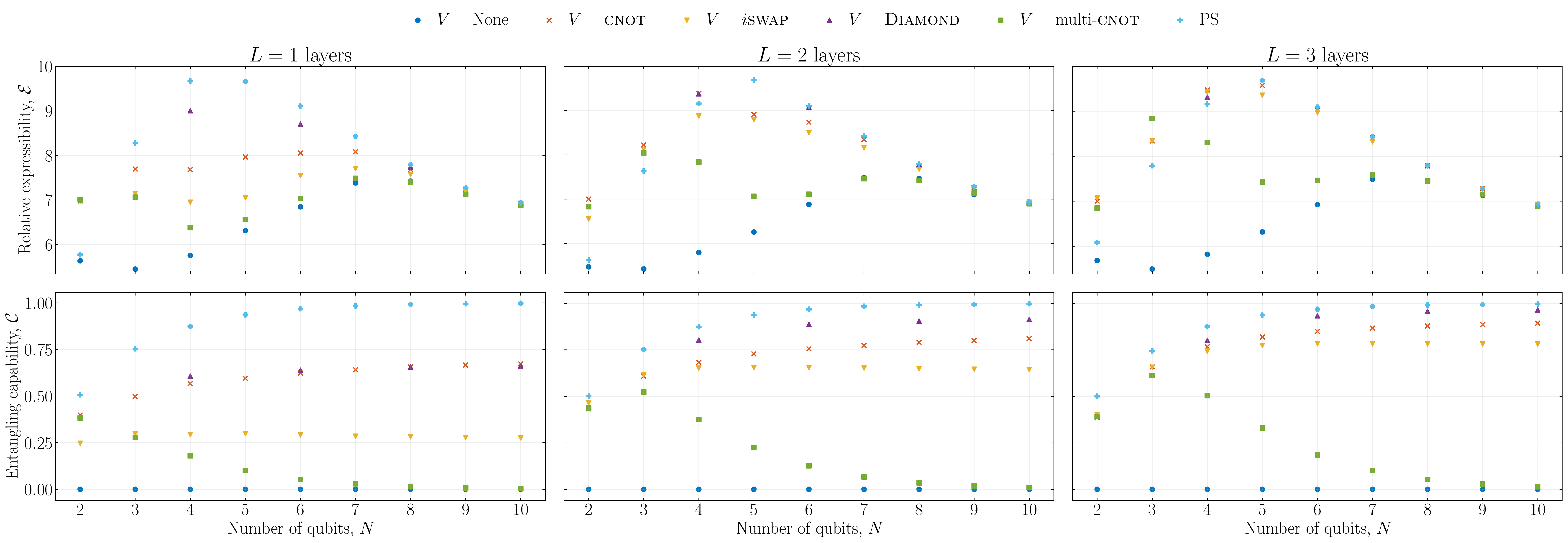}
	\caption{\emph{Top:} Relative expressibility and \emph{bottom:} entangling capability for different choices of $V$ gate as a function of the number of qubits. In order to ensure convergence all circuits are simulated with with the maximum number of rotations, i.e., $M$ rotations, with the exception of the photonics circuit which have zero single-qubit rotations.}
	\label{fig:qubitPlot}
\end{figure*}

In the light of the fact that all eight qubits PQC seems to converge approximately towards the same relative expressibility, we investigate the relative expressibility as a function of the number of qubits. We therefore plot the relative expressibility for different $V$ gates in the top part of Figure \ref{fig:qubitPlot}. We plot for up to three layers. In order to ensure that the relative expressibility has converged we use the maximum number of rotations, i.e., $M$ rotations. We use a selection of the same gates discussed in the previous section: No $V$-gate, \cnot{}s, \iswap{}s, diamond gates, the multi-qubit controlled \notgate gates, and a pulse shaper gate which is discussed in more detail in Section \ref{sec:photonics}.

For less than eight qubits we observe that the relative expressibility is quite scattered, with the entangling gates yielding the best relative expressibility, and the identity gate yielding worst relative expressibility. However, as the number of qubits increases the relative expressibility seems to converge towards $\mathcal{E} = 6.5 $ for all cases. 
This is because when the number of qubits increases the Haar measure peaks at low fidelity meaning that the lowest bins dominate in the calculation of the relative expressibility. When the number of qubit becomes sufficiently large only the lowest bin becomes relevant when calculating the relative expressibility. 
One should therefore consider increasing the number of bins if one wishes to investigate the relative expressibility for more qubits. However, in order to compare the relative expressibility for circuits with different numbers of qubits one should have the same number of bins, which is why we have chosen 75 bins for all calculations. This also means that even though the relative expressibility seems to converge towards a lower value for larger $N$ it does not necessarily mean these PQCs are worse than similar PQC for lower $N$. It is simply a result of the way the relative expressibility is defined. One should therefore be cautious when comparing across different number of qubits. 

Contrary to the relative expressibility the entangling capability is not dependent on the number of qubits, and we do not expect it to converge towards a specific value for all cases. We plot the entangling capability in the bottom part of Figure \ref{fig:qubitPlot}. Without any entangling gates the entangling capability is of course zero. When we introduce the \cnot gates we obtain an asymptotically increasing entangling capability. Not surprisingly, we see that more layers increase the entangling capability of the \cnot gates. We observe that for up to three layers there is a slow convergence towards one, i.e., 0.7 for one layer, 0.92 for two layers, and 0.95 for three layers.

The entangling capability of \iswap gates between each qubit is constant for a single layer around 0.3, which is significantly lower than for the \cnot gates, despite both having the maximum entangling power \cite{Williams2011}. Increasing the number of layers greatly increases the entangling capability of the circuit, however, it still remains inferior to a circuit with \cnot gate for the same amount of layers.
Turning to the diamond gate we see that for one layer its entangling capability is approximately equal to that of the \cnot gates, though it does seem to increase a bit slower than the \cnot gate. For two layers it outperforms the \cnot gates for three layers, and for three layers of the diamond gate it looks like the entangling capability converges towards one.
Contrary to the other entangling gates the entangling capability of the multi-qubit controlled \notgate gate converges towards zero as the number of qubits increases. This is probably because for a large number of control qubits the multi-qubit gate resembles the identity gate quite well, as it acts trivially on most of the qubits in the system.

\section{Photonic parameterized quantum circuits}\label{sec:photonics}

When considering PQC for HQC algorithms the focus is usually on the two-level qubits. However, there is nothing hindering us from considering higher dimensional quantum systems. In fact such high-dimensional systems arise naturally in quantum frequency comb photonic systems using fiber optics, where each photon may have more than two internal states \cite{Lukens2017,Lu2018,Reimer2019,Kues2017,Kues2019,Imany2019}. States featuring $d>2$ internal levels are typically denoted 'qudits'. 
Such photonic qudits have the advantage that their decoherence time is much longer than for other quantum computation schemes such as superconducting circuits. Qudits also have the advantage that due to their higher dimensions, less qudits are needed in order to create the same size Hilbert space compared to qubits. This could prove to be import in the current NISQ era, where a main struggle is qubit lifetimes, scaling the number of qubits, and multi-qubit gate fidelity.
One of the disadvantages of using qudits instead of qubits is one often has to find gate equivalences of qubit gates in qudit systems, e.g. how are Pauli gates mapped to photonic qudit gates, when Pauli gates are required \cite{Kottmann2020}. In this section we take a different approach. We do not try to find photonic gate equivalences to single-qubit rotation gates, instead we exploit the fact that in HQC algorithms we often simply need parameterized gates we can manipulate. Instead we employ native photonic qudit gates and vary these parameters in order to determine their relative expressibility and entangling capability. We note that even though this approach has advantages there are still disadvantages which should be tackled before qudits can be considered for quantum simulations. 
 
Quantum frequency comb photonics occurs naturally in high bandwidth laser and can be controlled by commercial off-the-shelve equipment, which easily can create superposition states, which in qubit encoding would be strongly entangled states. The major disadvantage of QFC photonics qudits is their weak ability to interact. We therefore consider a single qudit with $d=2^N$ states. By assuring that the number of states in the qudit is equal to some power of two, we can directly compare the circuit to a regular PQC with $N$ qubits by using a one-to-one mapping: $|1\rangle = |00\dots 00 \rangle$, $|2\rangle = |00 \dots 01\rangle$, $|3\rangle = |00 \dots 10 \rangle$, \dots, $|d\rangle = |11 \dots 11\rangle$.
As we consider a single qudit instead of $N$ qubits we cannot act on the state with usual single-qubit rotations. We are therefore forced to employ other gates to the qudit. The frequency comb qudit states $|\omega\rangle$ corresponds to a photon with the given frequency. Being in a frequency comb system, all neighboring qudit states have the same frequency difference $\Delta \omega$ and the states can thus be
numbered as $|n\rangle = |\omega_0 + n\Delta\omega\rangle$ with $|\omega_0\rangle$ being some offset frequency.
The qudit is then initialized in the even superposition state
\begin{equation}\label{eq:evensuperPos}
|i\rangle = \frac{1}{\sqrt{d}}\sum_{n=1}^{d} |n\rangle. 
\end{equation}
Note that if we employ our one-to-one mapping we obtain a highly entangled qubit state, from which we clearly see this schemes advantage when it comes to creating entangled states.

We first act on the qudit with a single pulse shaper (\ps) gate \cite{Monmayrant2010,Weiner2011}.
The pulse shaper gate is configured as a single Fourier-transform pulse shaper gate, which in principle works by separating each frequency and then independently and in parallel controlling the amplitude and phase of each frequency component. However, since amplitude modulation (attenuation) reduces the total population, we will only be using it for controlling the phases of each frequency. As stated, this can be done in parallel and independently, thus we can write the
corresponding gate operator as
\begin{equation}
\hat U_\text{PS} = \sum_{j=1}^{d} e^{i\theta_j} |j \rangle \langle j |,
\end{equation}
where $d$ is the number of frequency bins in the system, which should be equal to the number of qudit levels. The phase, $\theta_j$, of each mode is the parameter we wish to optimize in our PQC.
This means that there are $2^N$ parameters to optimize classically. Thus the number of parameters increases more rapidly than the case of regular PQCs as in Figure \ref{fig:circuitTemplate}, where the number of parameters increases linearly. This means that we increase the complexity of the classical part of the computation, however, it significantly decreases the quantum mechanical complexity of the calculation as we are now only concerned with a few gates per layer.

We calculate the relative expressibility and entangling capability of the pulse shaper gate acting on the state in Equation \ref{eq:evensuperPos}, i.e., $\hat U_\text{PS}|i\rangle$, for a different number of qudit levels. We plot the results in Figure \ref{fig:qubitPlot} together with qubit PQCs with $N$ qubits. This means that we are comparing a single $2^N$-dimensional qudit with $N$ qubits. We observe that the relative expressibility of the pulse shaper PQC matches the relative expressibility of the best qubit PQCs. 
We also see that the entangling capability of the pulse shaper gate beats all of the regular PQCs even for just a single layer. This fact means that we can significantly decrease the quantum mechanical part of the computation as we only require one gate in our circuit in order to obtain optimal relative expressibility and entangling capability.

\begin{figure*}
	\centering
	\includegraphics[width=\textwidth]{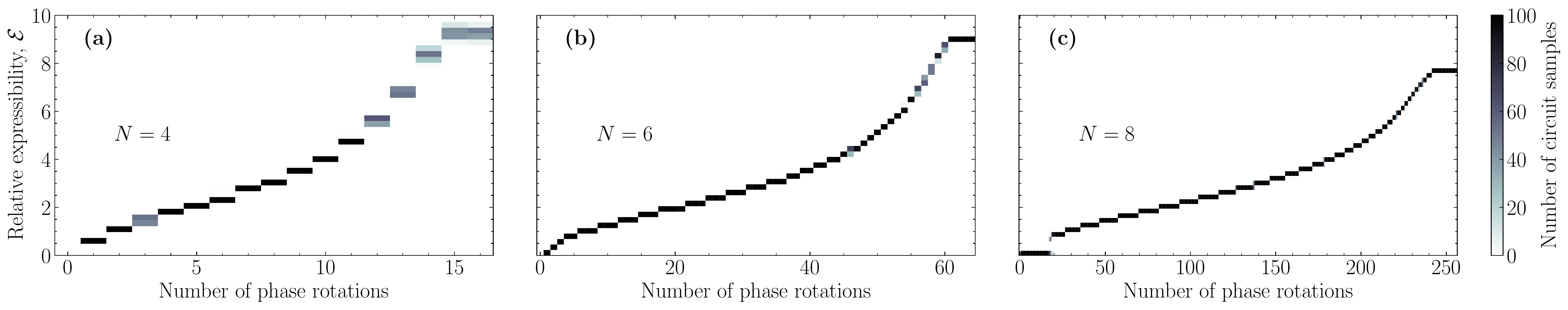}
	\caption{Relative expressibility of a \ps gate corresponding to $N=4, 6$, and 8 qubits as a function of the number of phase rotations in the gate.}
	\label{fig:exprPS}
\end{figure*}

However, there is an important caveat to the fact that the relative expressibility and entangling capability is superior for the \ps gate. The \ps gate employs $2^N$ phase rotations, which means just as many parameters. When $N$ increases $2^N$ becomes significantly larger than the number of parameters needed for the the PQC employing single-qubit rotations. In order to make a fair comparison between the number of parameters we investigate the relative expressibility of the \ps gate as a function of the number of phase rotations by sampling 100 random \ps gate for each number of rotations. We do not calculate the entangling capability as a function of the single-qubit rotations due to computational complexity. The result of the relative expressibility calculation can be seen in Figure \ref{fig:exprPS}. From this we see that it is only possible to save a few of the $2^N$ phase rotations, meaning that the \ps gate cannot compete with regular PQCs when it comes to reducing the number of parameters. This does, however, not mean that the \ps gate is useless in HQC algorithms, it could still be relevant in some algorithms were the number of parameters is less significant.

Despite large relative expressibility and entangling capability of the pulse shaper gate, it is still not evident that it would be efficient in HQC algorithms such as the variational quantum eigensolver and others. If we consider the resulting state when acting with the pulse shaper on the state in Equation \ref{eq:evensuperPos} we find
\begin{equation}
\hat U_\text{PS}|i\rangle = \frac{1}{\sqrt{d}} \sum_{j=1}^d e^{i\theta_j} |j\rangle.
\end{equation}
Thus we have obtained an even superposition with full control over the phases. However, it is not possible to remove any states from the superposition by varying the parameters. This could be a significant problem if the states are used for HQC algorithms. It is possible to vary the amplitudes of the \ps gate at the expense of decreasing the possibility of observing the photon. However, preliminary tests indicate that variation of amplitude in the PS gate does not lead to good results in an VQE algorithm.

\begin{figure}
	\includegraphics[scale=0.3]{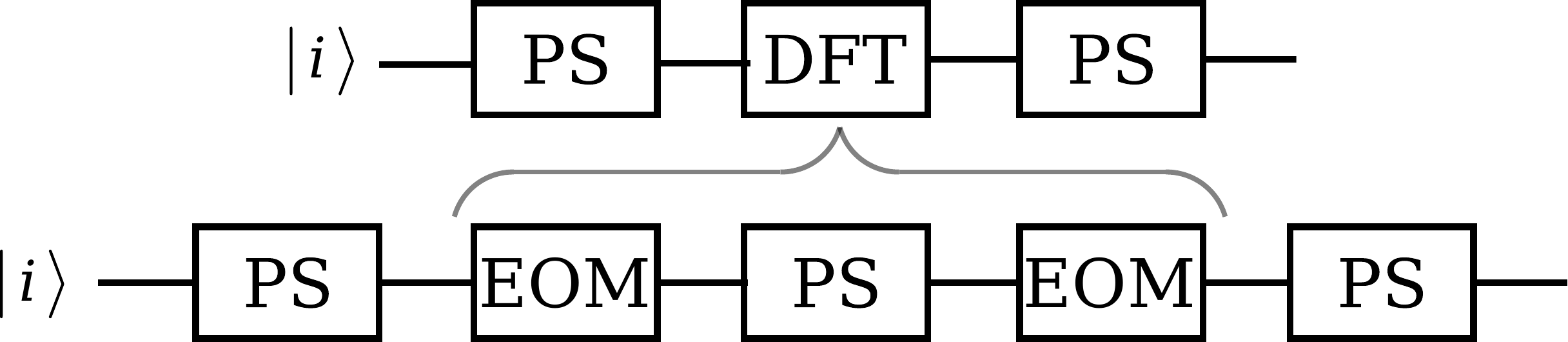}
	\caption{\emph{Upper:} Photonic frequency comb qudit PQC consisting of a pulse shaper gate (\ps) followed a discrete Fourier transform gate (\dft) followed by yet another pulse shaper gate. \emph{Lower:} The gate sequence used to realize the \dft gate consisting of a pulse shaper gate sandwiched in between two electro-optic modulators. The input state, $|i\rangle$, is an even superposition, see Equation \ref{eq:evensuperPos}.}
	\label{fig:dftps}
\end{figure}

Therefore we propose to add a discrete Fourier transform (\dft) gate \cite{Lu2018} followed by another \ps gate, see Figure \ref{fig:dftps}.
The Discrete Fourier Transform gate is essentially a uniform beamsplitter which splits the beam into a number of modes, where the phase of each mode is matched to that of the a discrete Fourier transform. Such a gate can be created by combining an electro-optic phase modulator (\eom), a pulse shaper, and then once more an \eom \cite{Lu2018}, see Figure \ref{fig:dftps}. 

The electro-optic phase modulator is composed of a waveguide which is formed by diffusing an electro-optic material (typically Ti) into a dielectric substrate (typically LiNbO$_3$) and a set of driving electrodes \cite{Capmany2011}. Consider the electrodes driven by a voltage function $\Delta V(t) = V_0 \cos(\Delta\omega_0 t + \theta)$ plus an unimportant constant contribution, which result in a global phase change that we can ignore. Classically this results in a phase shift of the incoming electromagnetic wave of $\exp(-i\pi\Delta V(t)/V_\pi)$. Here, $V_\pi$ is the DC voltage required for a constant $\pi$ phase shift. For simplicity we absorb the constants $\pi$ and $V_\pi$ into $\Delta V(t)$. The resulting quantum transformation of the electro-optic phase modulator in photon-frequency notation is then
\begin{equation}
\hat U_\text{EOM} = \int \frac{d\omega}{2\pi} \sum_{n=-V_0-1}^{V_0+1} (-ie^{-i\theta})^n J_n(V_0) \hat a^\dagger (\omega+ n\Delta\omega_0) \hat a(\omega),
\end{equation}
where $J_n$ are the Bessel functions of the first kind and $\hat a$ and $\hat a^\dagger$ are the photon annihilation and creation operators. Thus each \eom gate has two parameters $V_0$ and $\theta$ which can be controlled. Combining two \eom gate around a single pulse shaper gate can be used to realize a \dft gate \cite{Lu2018}, see Figure \ref{fig:dftps}.

Preliminary results show that that this combination of gates will perform well in an VQE algorithm utilizing frequency photonics systems, being able to find the ground state energy of H$_2$, LiH and BeH$_2$ with the same precision as the regular PQCs.
A calculation of the relative expressibility and entangling capability of the circuit in Figure \ref{fig:dftps} yields the same result as for the single pulse shaper gate, which we believe is the maximum possible relative expressibility and entangling capability for this choice of bins.

We have only considered systems consisting of a single qudit, however, it would be interesting to see if one could find some middle ground between one $2^N$-level qudit and $N$ qubits, where the number of parameters is optimal without considerably increasing the quantum mechanical complexity.

\section{Conclusion and outlook}\label{sec:conclusion}

We have investigated the relative expressibility and entangling capability for PQCs for a varying number of single-qubit rotations. We have that both the relative expressibility and entangling capability can be saturated using less single-qubit rotations than the maximum possible amount supported by the circuit. 
We have compared these metrics to actual VQE calculations of the ground state energy of small molecules using the same number of qubits. These results also show that the energy converges before reaching the maximum number of single-qubit rotations. In general the saturation point depends on the choice of entangling gate.
This could significantly decrease the classical computational complexity of many hybrid quantum classical algorithms. We also find that it is subordinate where these single-qubit rotations are placed in the circuit when the number of rotations is large. When the number of rotations is small one must spread the rotations evenly among the qubits, such that no qubit becomes saturated with rotations, in order to obtain an advantage.
We note that even though this saturation point is reached it does not mean that the given PQC is capable of finding any arbitrary state in the Hilbert space. It merely means that the given PQC has reached the limit of its capability.

Once the saturation point is achieved, the only remaining parameter to change is the type of multi-qubit gate in each layer. We find that highly entangling multi-qubit gates, such as the \cnot gate or the diamond gate \cite{Loft2019} reach the saturation point of the relative expressibility, $\mathcal{E}$ and entangling capability, $\mathcal{C}$, with less single-qubit rotations compared to less entangling gates.

It is our assessment that an efficient PQC can be created by employing a number of highly entangling gates such as the diamond gate or the \cnot in each layer, for three layers and then doing five to ten single-qubit rotations on each qubit. The placement of the single-qubit rotations can be randomized.

We have also considered if QFC photonics PQCs are suitable for HQC algorithms. Using a single pulse shaper gate, we obtain a relative expressibility and an entangling capability comparable to the best regular PQCs. We also used a QFC qudit to calculate the ground state of small molecules. This could open up for a new exploitation of QFC photonics in HQC algorithms, which could increase the computational power of such algorithms, since photonics might be able to accommodate more states than the current working number of qubits in e.g. superconducting circuits.

A high relative expressibility and entangling capability are of course not enough to prove that PQCs are useful for all HQC algorithms. A next step would then be to investigate the PQCs mentioned in this paper in HQC settings different from the VQE algorithm.

\medskip
\textbf{Acknowledgements} \par 
This work is supported by the Danish Council for Independent Research and the Carlsberg Foundation. 
The numerical results presented in this work were obtained at the Centre for Scientific Computing, Aarhus http://phys.au.dk/forskning/cscaa/. M.K. received support by the German Federal Ministry of Education and Research (Project PQuMAL).


\end{document}


\title{Supporting Information:\\ Reducing the amount of single-qubit rotations in VQE and related algorithms}

\author{S. E. Rasmussen$^{1,}$\thanks{stig@phys.au.dk}\phantom{*}, N. J. S. Loft$^{1}$, T. Bækkegaard$^{1}$, M. Kues$^{2}$, and N. T. Zinner$^{1,3,}$\thanks{zinner@phys.au.dk} \\
$^{1}$\textit{\affA} \\ $^{2}$\textit{Hannover Center for Optical Technologies (HOT), Leibniz University Hannover,}\\ \textit{D-30167 Hannover, Germany.} \\ $^{3}$\textit{\affB}}

\maketitle

\vspace{-1cm}
\section{Diamond gate}\label{app:diamond}

Here we present the Hamiltonian and unitary of the Diamond gate. For at more detailed discussion see \cite{Loft2019}. The Hamiltonian of the diamond gate is
\begin{equation}
H= -\frac{1}{2} (\Omega + \Delta) (\sigma_z^{T1} + \sigma_z^{T2}) - \frac{1}{2}\Omega (\sigma_z^{C1} + \sigma_z^{C2}) + J_C \sigma_y^{C1} \sigma_y^{C2} + J(\sigma_y^{C1} + \sigma_y^{C2})(\sigma_y^{T1} + \sigma_y^{T2}).
\end{equation}
In the main text we set $J_C = 0$ and obtain the following unitary of the four qubit gate

\begin{equation}
\setcounter{MaxMatrixCols}{20}
U_\text{dia} = \begin{pmatrix}
\phantom{-}1 & 0 & 0 & 0 & 0 & 0 & 0 & 0 & 0 & 0 & 0 & 0 & 0 & 0 & 0 & 0\\
0 & 0 &-1 & 0 & 0 & 0 & 0 & 0 & 0 & 0 & 0 & 0 & 0 & 0 & 0 & 0\\
0 &-1 & 0 & 0 & 0 & 0 & 0 & 0 & 0 & 0 & 0 & 0 & 0 & 0 & 0 & 0\\
0 & 0 & 0 &-1 & 0 & 0 & 0 & 0 & 0 & 0 & 0 & 0 & 0 & 0 & 0 & 0\\
0 & 0 & 0 & 0 & 0 & 0 & 0 & 0 &-1 & 0 & 0 & 0 & 0 & 0 & 0 & 0\\
0 & 0 & 0 & 0 & 0 & 1 & 0 & 0 & 0 & 0 & 0 & 0 & 0 & 0 & 0 & 0\\
0 & 0 & 0 & 0 & 0 & 0 & 1 & 0 & 0 & 0 & 0 & 0 & 0 & 0 & 0 & 0\\
0 & 0 & 0 & 0 & 0 & 0 & 0 & 0 & 0 & 0 & 0 &-1 & 0 & 0 & 0 & 0\\
0 & 0 & 0 & 0 &-1 & 0 & 0 & 0 & 0 & 0 & 0 & 0 & 0 & 0 & 0 & 0\\
0 & 0 & 0 & 0 & 0 & 0 & 0 & 0 & 0 & 1 & 0 & 0 & 0 & 0 & 0 & 0\\
0 & 0 & 0 & 0 & 0 & 0 & 0 & 0 & 0 & 0 & 1 & 0 & 0 & 0 & 0 & 0\\
0 & 0 & 0 & 0 & 0 & 0 & 0 &-1 & 0 & 0 & 0 & 0 & 0 & 0 & 0 & 0\\
0 & 0 & 0 & 0 & 0 & 0 & 0 & 0 & 0 & 0 & 0 & 0 &-1 & 0 & 0 & 0\\
0 & 0 & 0 & 0 & 0 & 0 & 0 & 0 & 0 & 0 & 0 & 0 & 0 & 0 &-1 & 0\\
0 & 0 & 0 & 0 & 0 & 0 & 0 & 0 & 0 & 0 & 0 & 0 & 0 &-1 & 0 & 0\\
0 & 0 & 0 & 0 & 0 & 0 & 0 & 0 & 0 & 0 & 0 & 0 & 0 & 0 & 0 & 1
\end{pmatrix},
\end{equation}
which act on the state $|C_1 C_2 T_1 T_2 \rangle$.

An example of a decomposition of the Diamond gate into standard gates is as follows

\begin{equation*}
	\Qcircuit @C=1.4em @R=1em {
		\lstick{T_1} & \targ     & \ctrl{1}   & \targ     & \qw & \qw & \qw & \qw & \qw & \qw & \qw & \ctrl{2} & \qw & \ctrl{2} & \gate{Z} & \targ & \ctrl{1} & \targ & \qw \\
		\lstick{T_2} & \ctrl{-1} & \gate{H}   & \ctrl{-1} & \qw & \ctrl{1} & \qw & \ctrl{1} & \ctrl{1} & \ctrl{1} & \qw & \qw & \qw & \qw & \qw & \ctrl{-1} & \gate{H} & \ctrl{-1} & \qw\\
		\lstick{C_1} & \ctrl{1}  & \qswap     & \gate{Z} & \qw & \ctrl{1} & \qw & \qswap & \ctrl{-1} & \qw & \qw & \ctrl{1} & \qw & \qswap & \qw & \qw & \qw & \qw & \qw  \\
		\lstick{C_2} & \ctrl{-1} & \qswap\qwx & \gate{Z} & \gate{H} & \targ & \gate{H} & \qswap\qwx & \qw & \ctrl{-1} & \gate{H} & \targ & \gate{H} & \qswap\qwx & \qw & \qw & \qw & \qw & \qw 
	}
\end{equation*}

Using the open-source Python toolbox \textsc{Qiskit} \cite{qiskit} we find that the gate can be decomposed into 42 \cnot gates and 49 single-qubit rotations.

\clearpage
\section{Construction of the chemical Hamiltonians used for the VQE algorithm}

\begin{table}
	\centering
	\caption{Overview of the three different molecules we consider in the VQE algorithm. Molecule bond length and angles are found at Ref. \cite{cccbdb2019}.}
	\label{tab:molecules}
	\begin{tabular}{lccc}
		& LiH & BeH$_2$ & OH \\
		\hline
		Bond length [Angstrom] & 1.595 & 1.334 & 0.964 \\
		Angle [degrees] & - & 180 & - \\
		Core orbitals & 1 & 1 & 1 \\
		Active space & 2 & 3 & 4 \\
		Qubits & 4 & 6 & 8 \\
		\hline
	\end{tabular}
\end{table}

We consider three different molecules for our VQE calculations in order to compare with the expressibility and entangling capability plots. The three molecules we consider are LiH, BeH$_2$, and OH, which we all consider at their equilibrium bond length (and angle in the case of BeH$_2$). For each molecule we consider only the active space of the molecular orbitals, disregarding both core orbitals and virtual orbitals. We first calculate the full molecular Hamiltonian using PySCF \cite{PySCF} and OpenFermion \cite{openFermion2017}. Then by means of Bravyi-Kitaev mapping we arrive at the reduced Hamiltonian on the form
\begin{equation}
\hat H = \sum_{\alpha_n\in X, Y, Z, I} h_{\alpha} \prod_{n=1}^{N}\sigma_n^{\alpha_n},
\end{equation}
where $\sigma_{n}^{X,Y,Z}$ are the Pauli operators and $\sigma_I^n$ is the identity on the $n$th qubit out of a total of $N$ qubits. The coefficients $h_\alpha$ is given by the Bravyi-Kitaev mapping. The sum is over all $N$ permutations of the Pauli operators and the identity. The number of qubits in the Hamiltonian is twice the active space of the given molecule. See Table \ref{tab:molecules} for data on each molecule. In Table \ref{tab:LiH}-\ref{tab:OH} we present the $h_\alpha$-coefficients for each molecule.

\begin{table}
	\centering
	\caption{Coefficients in the qubit Hamiltonian for LiH at bond length $\SI{1.595}{\angstrom}$.}
	\label{tab:LiH}
	\begin{tabular}{llllll}
		Operators\qquad & $h_\alpha$ (Hartree)\hspace{0.5cm} & Operators\qquad & $h_\alpha$ (Hartree)\hspace{0.5cm} & Operators\qquad & $h_\alpha$ (Hartree) \\
		\hline
		$IIII$ & $-7.508666$  & $IIZI$ & $0.156354$  & $IXII$ & $0.013941$  \\
		$IXIZ$ & $-0.013941$  & $IXZI$ & $0.156354$  & $IZII$ & $-0.013941$  \\
		$IZIZ$ & $0.013941$  & $IZZZ$ & $-0.014942$  & $XXXI$ & $-0.014942$  \\
		$XXXZ$ & $0.122001$  & $XYYI$ & $0.012103$  & $XZXI$ & $0.012103$  \\
		$XZXZ$ & $-0.012103$  & $YXYI$ & $0.012103$  & $YXYZ$ & $0.003241$  \\
		$YYXI$ & $0.003241$  & $YZYI$ & $0.003241$  & $YZYZ$ & $0.003241$  \\
		$ZIII$ & $0.052733$  & $ZIZI$ & $0.055974$  & $ZIZZ$ & $0.001838$  \\
		$ZXIZ$ & $0.001838$  & $ZXZI$ & $0.055974$  & $ZXZZ$ & $0.001838$  \\
		$ZZII$ & $-0.001838$  & $ZZZI$ & $0.052733$  & $ZZZZ$ & $0.084497$  \\
		\hline
	\end{tabular}
\end{table}

\begin{table}
	\centering
	\caption{Coefficients in the qubit Hamiltonian for BeH$_2$ at bond length $\SI{1.334}{\angstrom}$.}
	\label{tab:BeH2}
	\begin{tabular}{llllll}
		Operators\qquad & $h_\alpha$ (Hartree)\hspace{0.5cm} & Operators\qquad & $h_\alpha$ (Hartree)\hspace{0.5cm} & Operators\qquad & $h_\alpha$ (Hartree) \\
		\hline
		$IIIIII$ & $-14.512197$  & $IIIIIZ$ & $0.321081$  & $IIIIZI$ & $0.321081$  \\
		$IIIIZZ$ & $0.309786$  & $IIXIXZ$ & $0.309786$  & $IIYIYZ$ & $0.128535$  \\
		$IIZIII$ & $0.128535$  & $IIZIZI$ & $0.099645$  & $IIZIZZ$ & $0.041110$  \\
		$IZIIII$ & $0.041110$  & $IZIZII$ & $0.041110$  & $IZXZXI$ & $0.041110$  \\
		$IZYZYI$ & $0.012360$  & $IZZZII$ & $0.012360$  & $IZZZZI$ & $0.012360$  \\
		$IZZZZZ$ & $0.012360$  & $XIIIXZ$ & $0.061871$  & $XZIIXI$ & $0.102981$  \\
		$XZXIII$ & $0.079954$  & $XZXZII$ & $0.092313$  & $YIIIYZ$ & $0.102981$  \\
		$YZIIYI$ & $0.061871$  & $YZYIII$ & $0.092313$  & $YZYZII$ & $0.079954$  \\
		$ZIIIII$ & $0.108772$  & $ZIIIZI$ & $0.003666$  & $ZIIIZZ$ & $0.003666$  \\
		$ZIZIII$ & $0.003666$  & $ZIZZII$ & $0.003666$  & $ZZIIII$ & $0.085489$  \\
		$ZZIIZI$ & $0.089155$  & $ZZIIZZ$ & $0.089155$  & $ZZZIII$ & $0.085489$  \\
		\hline
	\end{tabular}
\end{table}

\begin{table}
	\centering
	\caption{Coefficients in the qubit Hamiltonian for OH at bond length $\SI{0.964}{\angstrom}$.}
	\label{tab:OH}
	\begin{tabular}{llllll}
		Operators\qquad & $h_\alpha$ (Hartree)\hspace{0.5cm} & Operators\qquad & $h_\alpha$ (Hartree)\hspace{0.5cm} & Operators\qquad & $h_\alpha$ (Hartree) \\
		\hline
		$IIIIIIII$ & $-69.450312$  & $IIIIIIZI$ & $1.440827$  & $IIIIIZII$ & $0.036733$  \\
		$IIIIXZXI$ & $-0.036733$  & $IIIIYZYI$ & $1.440827$  & $IIIIZIII$ & $-0.036733$  \\
		$IIIIZIZI$ & $0.036733$  & $IIIIZZII$ & $1.078773$  & $IIIIZZZI$ & $1.078773$  \\
		$IIIZIZIZ$ & $1.097665$  & $IIIZIZZZ$ & $1.097665$  & $IIIZXZXZ$ & $1.097665$  \\
		$IIIZYZYZ$ & $1.097665$  & $IIIZZIZZ$ & $0.188447$  & $IIIZZZZZ$ & $-0.001082$  \\
		$IIXIXZII$ & $-0.001082$  & $IIXZIZXZ$ & $0.001082$  & $IIYIYZII$ & $-0.001082$  \\
		$IIYZIZYZ$ & $0.035778$  & $IIZIIIII$ & $0.035778$  & $IIZIIIZI$ & $0.035778$  \\
		$IIZIZIII$ & $0.035778$  & $IIZIZZII$ & $0.037926$  & $IIZZIZZZ$ & $0.037926$  \\
		$IXIIIIII$ & $0.037926$  & $IXIZIIII$ & $0.037926$  & $IXXIIIXI$ & $0.037926$  \\
		$IXXIXIII$ & $0.037926$  & $IXYIIIYI$ & $0.037926$  & $IXYIYIII$ & $0.037926$  \\
		$IXZIIIII$ & $0.133454$  & $IXZIIIZI$ & $0.169232$  & $IXZIZIII$ & $-0.013167$  \\
		$IXZIZZII$ & $-0.013167$  & $IXZZIZZZ$ & $-0.011614$  & $IZIIIIII$ & $-0.011614$  \\
		$IZIZIIII$ & $0.011614$  & $IZXZIIXI$ & $-0.011614$  & $IZXZXIII$ & $-0.011614$  \\
		$IZYZIIYI$ & $-0.011614$  & $IZYZYIII$ & $0.011614$  & $IZZIIZZZ$ & $-0.011614$  \\
		$IZZZIIII$ & $0.152838$  & $IZZZIIZI$ & $0.006517$  & $IZZZZIII$ & $-0.006517$  \\
		$IZZZZZII$ & $0.190764$  & $XIIIXZII$ & $0.018131$  & $XIIZIZXZ$ & $-0.018131$  \\
		$XXIZIIXI$ & $0.152838$  & $XXIZXIII$ & $0.006517$  & $XXXIIIII$ & $-0.006517$  \\
		$XXXZIIII$ & $0.190764$  & $XYYIIIII$ & $0.018131$  & $XYYIIIZI$ & $-0.018131$  \\
		$XYYIZIII$ & $0.169232$  & $XYYIZZII$ & $-0.013167$  & $XYYZIZZZ$ & $0.013167$  \\
		$XYZIYZII$ & $0.011614$  & $XYZZIZYZ$ & $-0.011614$  & $XZIIIIXI$ & $0.011614$  \\
		$XZIIXIII$ & $0.011614$  & $XZXIIIII$ & $0.011614$  & $XZXZIIII$ & $-0.011614$  \\
		$YIIIYZII$ & $0.011614$  & $YIIZIZYZ$ & $0.011614$  & $YXIZIIYI$ & $0.133454$  \\
		$YXIZYIII$ & $0.190764$  & $YXYIIIII$ & $-0.018131$  & $YXYZIIII$ & $0.018131$  \\
		$YYXIIIII$ & $0.152838$  & $YYXIIIZI$ & $-0.006517$  & $YYXIZIII$ & $0.006517$  \\
		$YYXIZZII$ & $0.190764$  & $YYXZIZZZ$ & $-0.018131$  & $YYZIXZII$ & $0.018131$  \\
		$YYZZIZXZ$ & $0.152838$  & $YZIIIIYI$ & $-0.006517$  & $YZIIYIII$ & $0.006517$  \\
		$YZYIIIII$ & $0.179377$  & $YZYZIIII$ & $0.012408$  & $ZIIIIIII$ & $0.012408$  \\
		$ZIIIIIZI$ & $0.012408$  & $ZIIIZIII$ & $0.012408$  & $ZIIIZZII$ & $0.012408$  \\
		$ZIIZIZZZ$ & $0.012408$  & $ZIZIIIII$ & $0.012408$  & $ZIZZIIII$ & $0.012408$  \\
		$ZXIIIZZZ$ & $0.159809$  & $ZXIZIIII$ & $0.172217$  & $ZXIZIIZI$ & $0.159809$  \\
		$ZXIZZIII$ & $0.172217$  & $ZXIZZZII$ & $0.172217$  & $ZXZIIIII$ & $0.159809$  \\
		$ZXZZIIII$ & $0.172217$  & $ZYXIYZII$ & $0.159809$  & $ZYXZIZYZ$ & $0.220040$  \\
		$ZYYIXZII$ & $0.011861$  & $ZYYZIZXZ$ & $0.011861$  & $ZZIIIIII$ & $0.011861$  \\
		$ZZIIIIZI$ & $0.011861$  & $ZZIIZIII$ & $0.184456$  & $ZZIIZZII$ & $0.196318$  \\
		$ZZIZIZZZ$ & $0.196318$  & $ZZZIIIII$ & $0.184456$  & $ZZZZIIII$ & $0.220040$  \\
		
		\hline
	\end{tabular}
\end{table}

\clearpage
\section{Expressibility, entangling capability, and VQE plots}\label{app:exprPlots}

Here we present expressibility, entangling capability, and VQE plots for layers $L=1,2,3$, and 4 for $N=4,6$, and 8 qubits.

\begin{figure}[h]
	\centering
	\includegraphics[width=\textwidth]{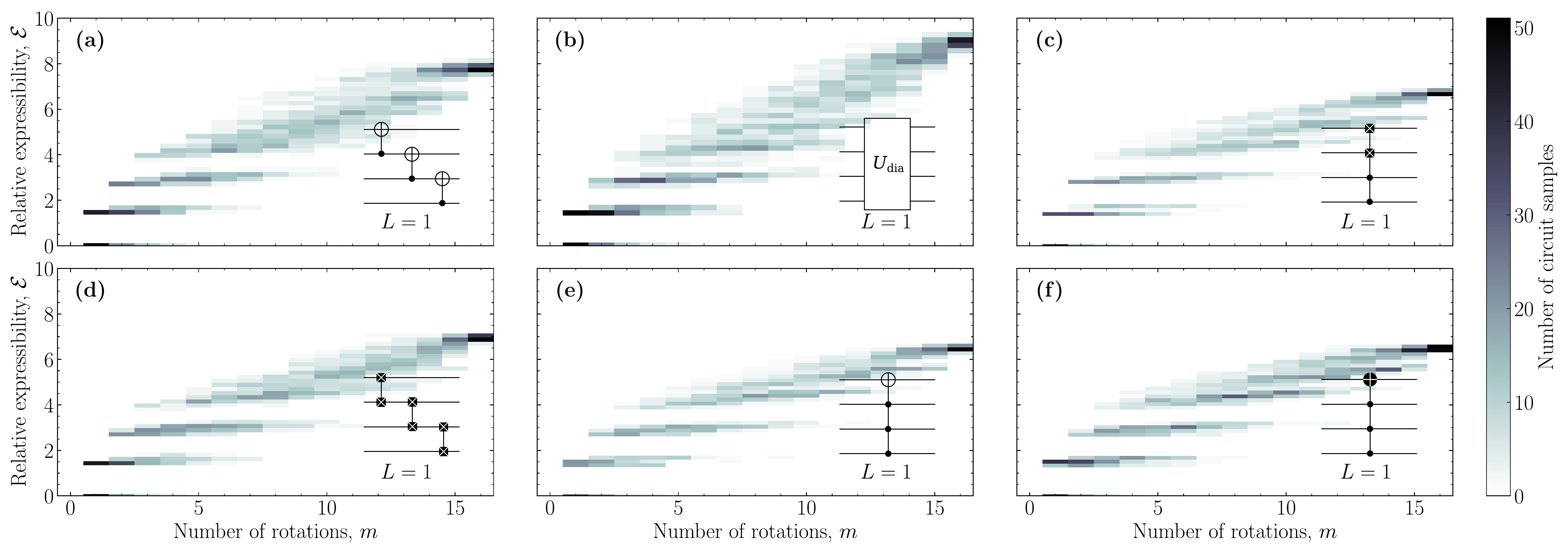}
	\caption{\emph{Relative expressibility} of different circuits with $N=4$ qubits and $L=1$ layer as a function of the number of rotations. The inset in the right upper corner of each plot shows the entangling gate, $V$, used in each layer. \textbf{(a)} Three \cnot{}s, \textbf{(b)} diamond gate, \textbf{(c)} double controlled \iswap, \textbf{(d)} three \iswap{}s, \textbf{(e)} triple controlled \notgate, \textbf{(f)} triple controlled $i$\notgate.}
	\label{fig:exprL1}
\end{figure}

\begin{figure}
	\centering
	\includegraphics[width=\textwidth]{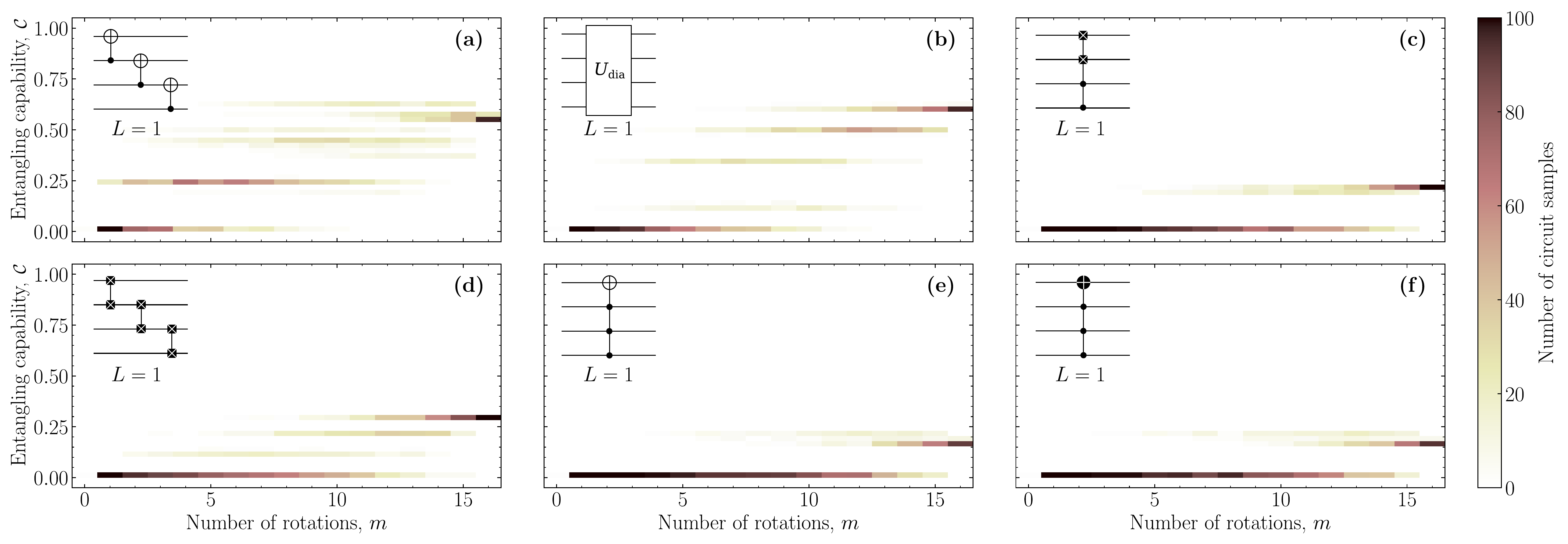}
	\caption{\emph{Entangling capability} of different circuits with $N=4$ qubits and $L=1$ layer as a function of the number of rotations. The inset in the right upper corner of each plot shows the entangling gate, $V$, used in each layer. \textbf{(a)} Three \cnot{}s, \textbf{(b)} diamond gate, \textbf{(c)} double controlled \iswap, \textbf{(d)} three \iswap{}s, \textbf{(e)} triple controlled \notgate, \textbf{(f)} triple controlled $i$\notgate.}
	\label{fig:entL1}
\end{figure}

\begin{figure}
	\includegraphics[width=\textwidth]{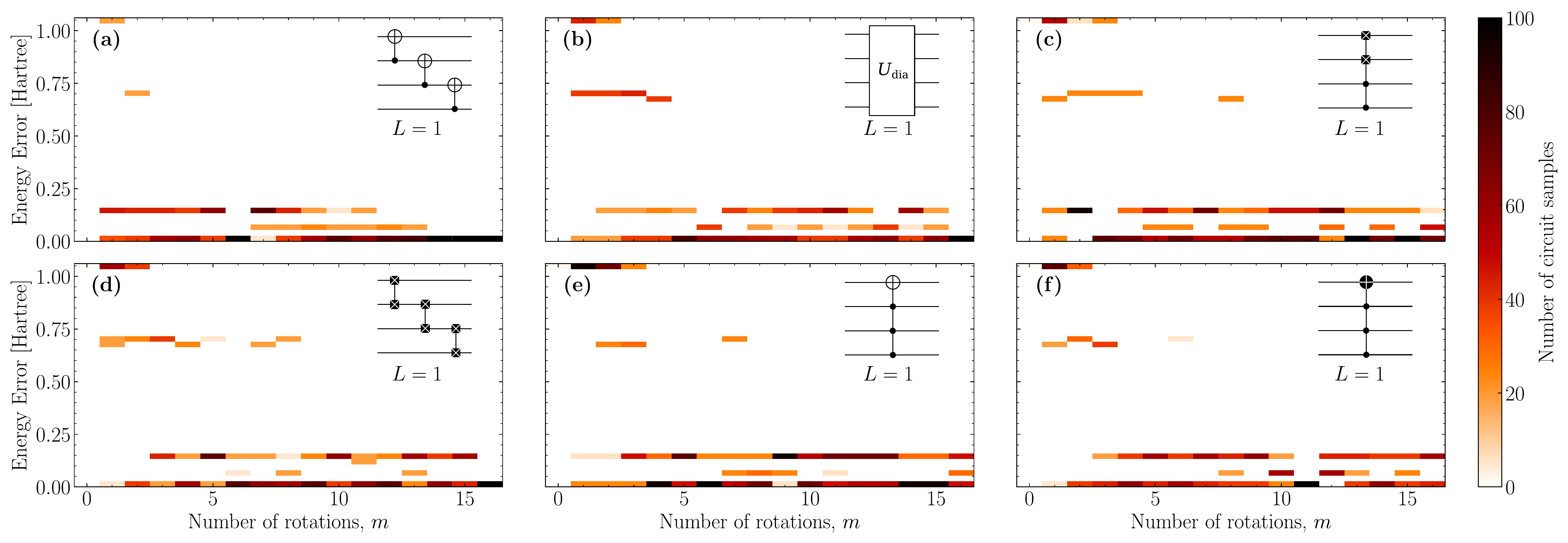}
	\caption{\emph{Energy error} of the ground state LiH found using VQE of different circuits with $N=4$ qubits and $L=1$ layer as a function of the number of rotations. The energy found using VQE is compared to a classical diagonalization of the reduced Hamiltonian, which yields the energy error. The inset in the right upper corner of each plot shows the entangling gate, $V$, used in each layer. \textbf{(a)} Three \cnot{}s, \textbf{(b)} diamond gate, \textbf{(c)} double controlled \iswap, \textbf{(d)} three \iswap{}s, \textbf{(e)} triple controlled \notgate, \textbf{(f)} triple controlled $i$\notgate.}
	\label{fig:vqeL1}
\end{figure}

\begin{figure}
	\centering
	\includegraphics[width=\textwidth]{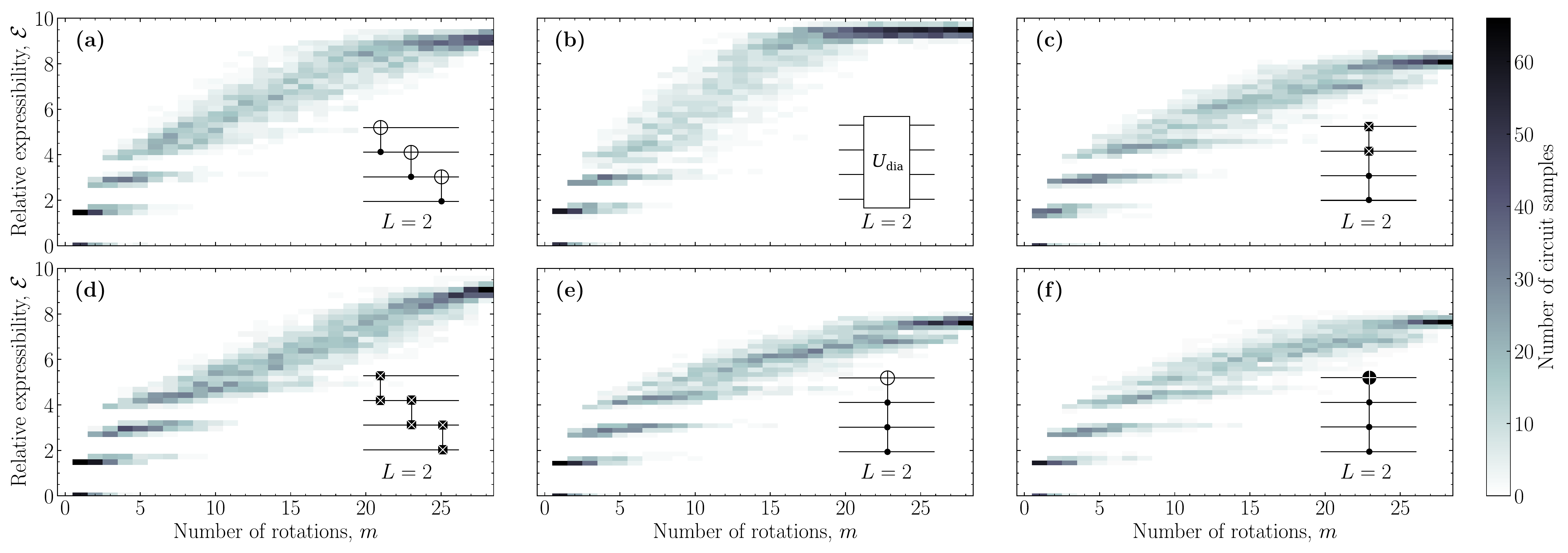}
	\caption{\emph{Relative expressibility} of different circuits with $N=4$ qubits and $L=2$ layers as a function of the number of rotations. The inset in the right upper corner of each plot shows the entangling gate, $V$, used in each layer. \textbf{(a)} Three \cnot{}s, \textbf{(b)} diamond gate, \textbf{(c)} double controlled \iswap, \textbf{(d)} three \iswap{}s, \textbf{(e)} triple controlled \notgate, \textbf{(f)} triple controlled $i$\notgate.}
	\label{fig:exprL2}
\end{figure}

\begin{figure}
	\centering
	\includegraphics[width=\textwidth]{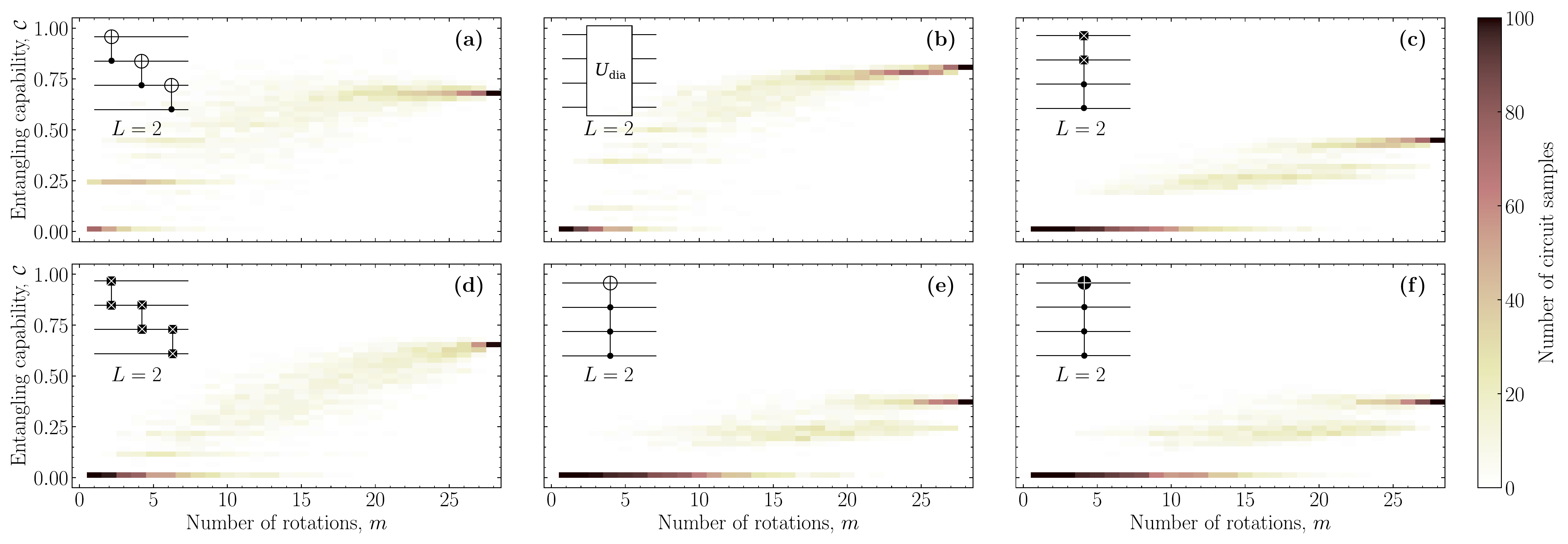}
	\caption{\emph{Entangling capability} of different circuits with $N=4$ qubits and $L=2$ layers as a function of the number of rotations. The inset in the right upper corner of each plot shows the entangling gate, $V$, used in each layer. \textbf{(a)} Three \cnot{}s, \textbf{(b)} diamond gate, \textbf{(c)} double controlled \iswap, \textbf{(d)} three \iswap{}s, \textbf{(e)} triple controlled \notgate, \textbf{(f)} triple controlled $i$\notgate.}
	\label{fig:entL2}
\end{figure}

\begin{figure}
	\includegraphics[width=\textwidth]{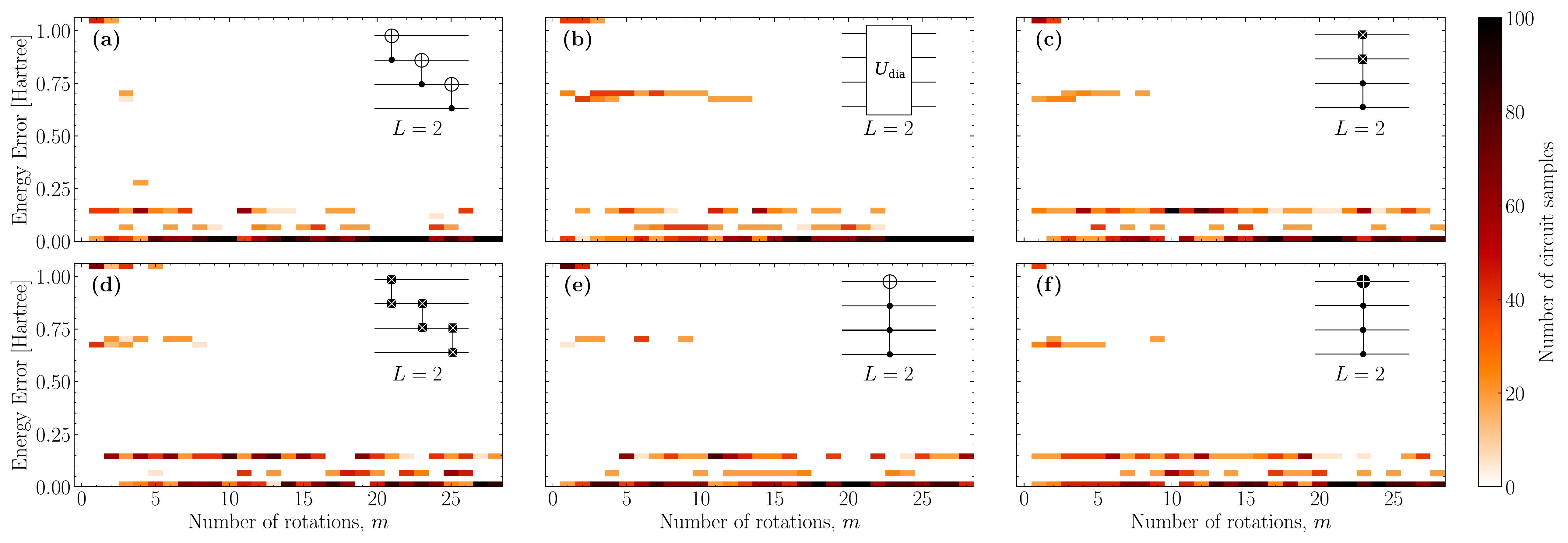}
	\caption{\emph{Energy error} of the ground state LiH found using VQE of different circuits with $N=4$ qubits and $L=2$ layers as a function of the number of rotations. The energy found using VQE is compared to a classical diagonalization of the reduced Hamiltonian, which yields the energy error. The inset in the right upper corner of each plot shows the entangling gate, $V$, used in each layer. \textbf{(a)} Three \cnot{}s, \textbf{(b)} diamond gate, \textbf{(c)} double controlled \iswap, \textbf{(d)} three \iswap{}s, \textbf{(e)} triple controlled \notgate, \textbf{(f)} triple controlled $i$\notgate.}
	\label{fig:vqeL2}
\end{figure}

\begin{figure}
	\centering
	\includegraphics[width=\textwidth]{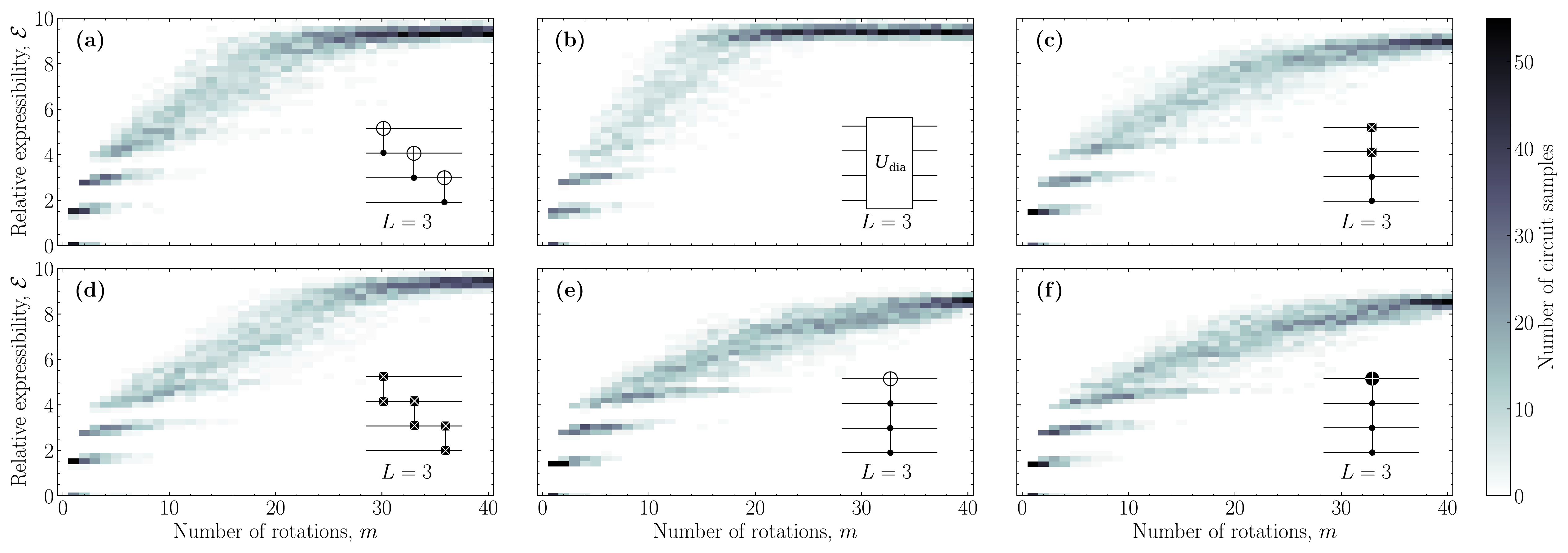}
	\caption{\emph{Relative expressibility} of different circuits with $N=4$ qubits and $L=3$ layers as a function of the number of rotations. The inset in the right upper corner of each plot shows the entangling gate, $V$, used in each layer. \textbf{(a)} Three \cnot{}s, \textbf{(b)} diamond gate, \textbf{(c)} double controlled \iswap, \textbf{(d)} three \iswap{}s, \textbf{(e)} triple controlled \notgate, \textbf{(f)} triple controlled $i$\notgate.}
	\label{fig:exprL3}
\end{figure}

\begin{figure}
	\centering
	\includegraphics[width=\textwidth]{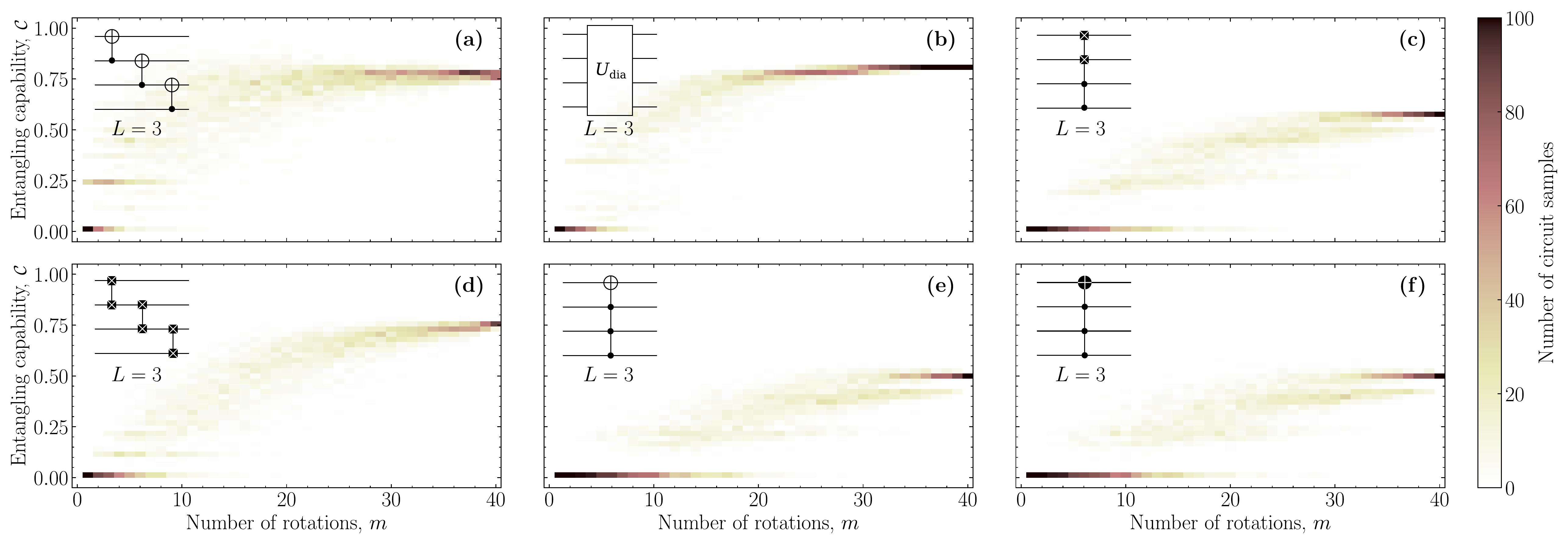}
	\caption{\emph{Entangling capability} of different circuits with $N=4$ qubits and $L=3$ layers as a function of the number of rotations. The inset in the right upper corner of each plot shows the entangling gate, $V$, used in each layer. \textbf{(a)} Three \cnot{}s, \textbf{(b)} diamond gate, \textbf{(c)} double controlled \iswap, \textbf{(d)} three \iswap{}s, \textbf{(e)} triple controlled \notgate, \textbf{(f)} triple controlled $i$\notgate.}
	\label{fig:entL3}
\end{figure}

\begin{figure}
	\includegraphics[width=\textwidth]{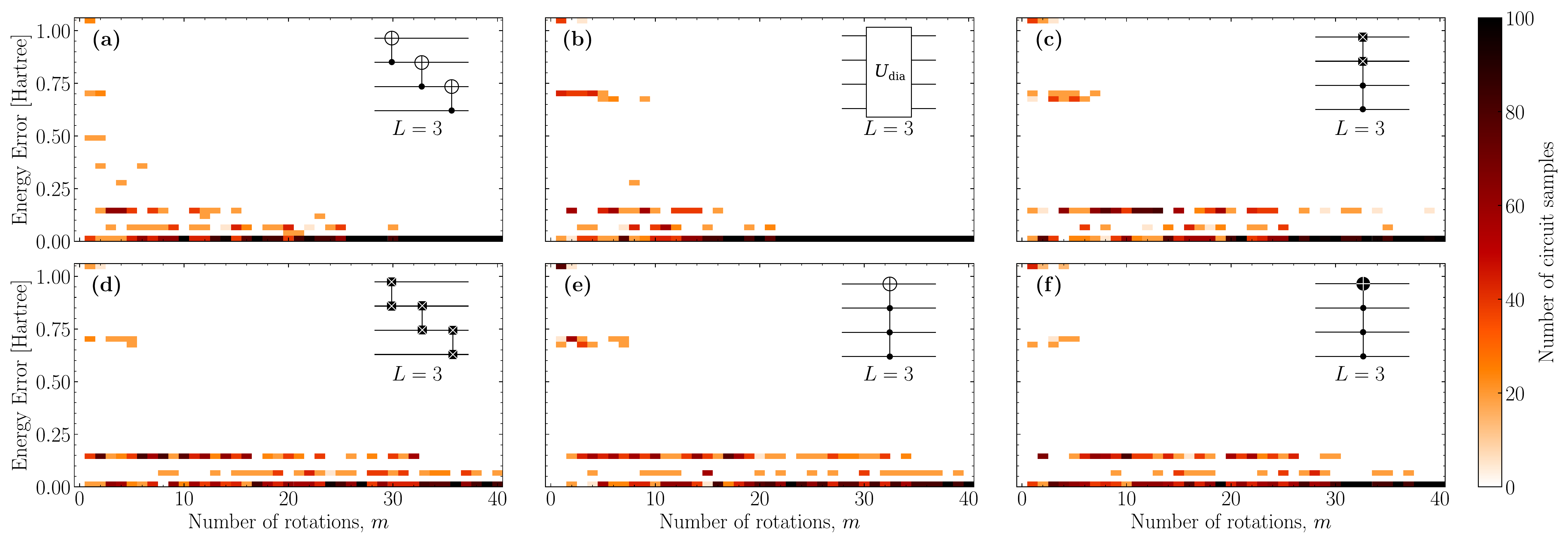}
	\caption{\emph{Energy error} of the ground state LiH found using VQE of different circuits with $N=4$ qubits and $L=3$ layers as a function of the number of rotations. The energy found using VQE is compared to a classical diagonalization of the reduced Hamiltonian, which yields the energy error. The inset in the right upper corner of each plot shows the entangling gate, $V$, used in each layer. \textbf{(a)} Three \cnot{}s, \textbf{(b)} diamond gate, \textbf{(c)} double controlled \iswap, \textbf{(d)} three \iswap{}s, \textbf{(e)} triple controlled \notgate, \textbf{(f)} triple controlled $i$\notgate.}
	\label{fig:vqeL3}
\end{figure}

\begin{figure}
	\centering
	\includegraphics[width=\textwidth]{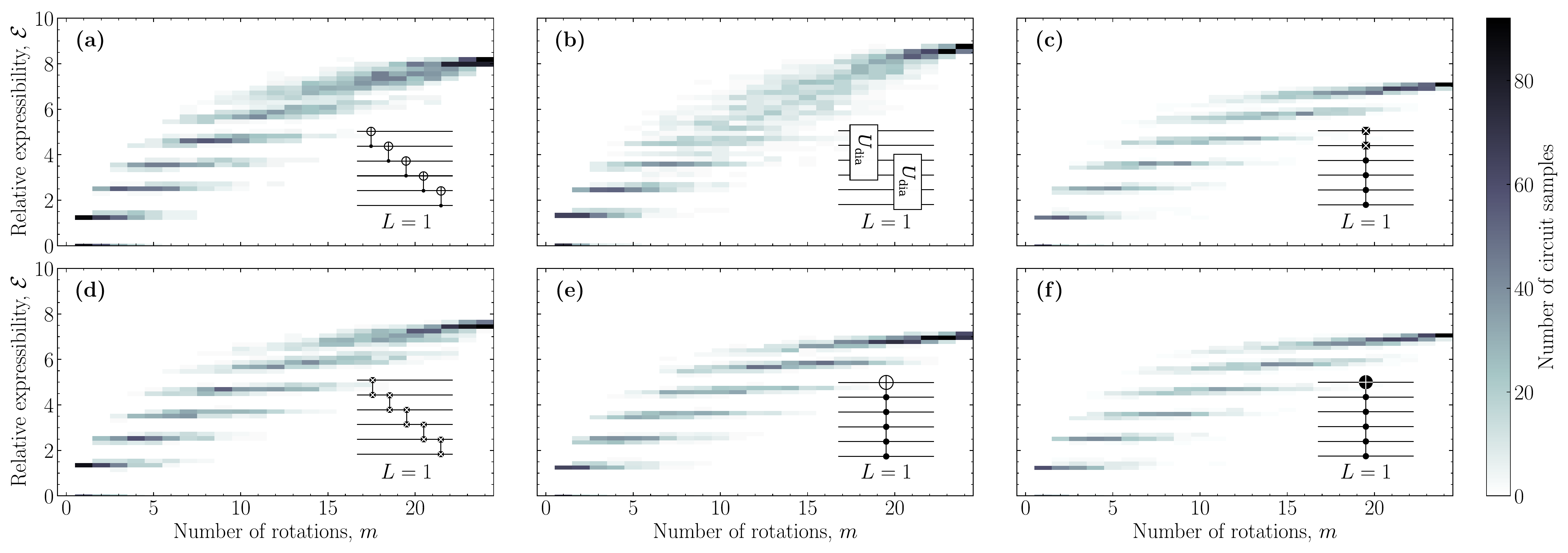}
	\caption{\emph{Relative expressibility} of different circuits with $N=6$ qubits and $L=1$ layer as a function of the number of rotations. The inset in the right upper corner of each plot shows the entangling gate, $V$, used in each layer. \textbf{(a)} Five \cnot{}s, \textbf{(b)} two diamond gates, \textbf{(c)} \iswap with four control qubits, \textbf{(d)} five \iswap{}s, \textbf{(e)} \notgate with five control qubits, \textbf{(f)} $i$\notgate with five control qubits.}
	\label{fig:exprL1_q6}
\end{figure}

\begin{figure}
	\centering
	\includegraphics[width=\textwidth]{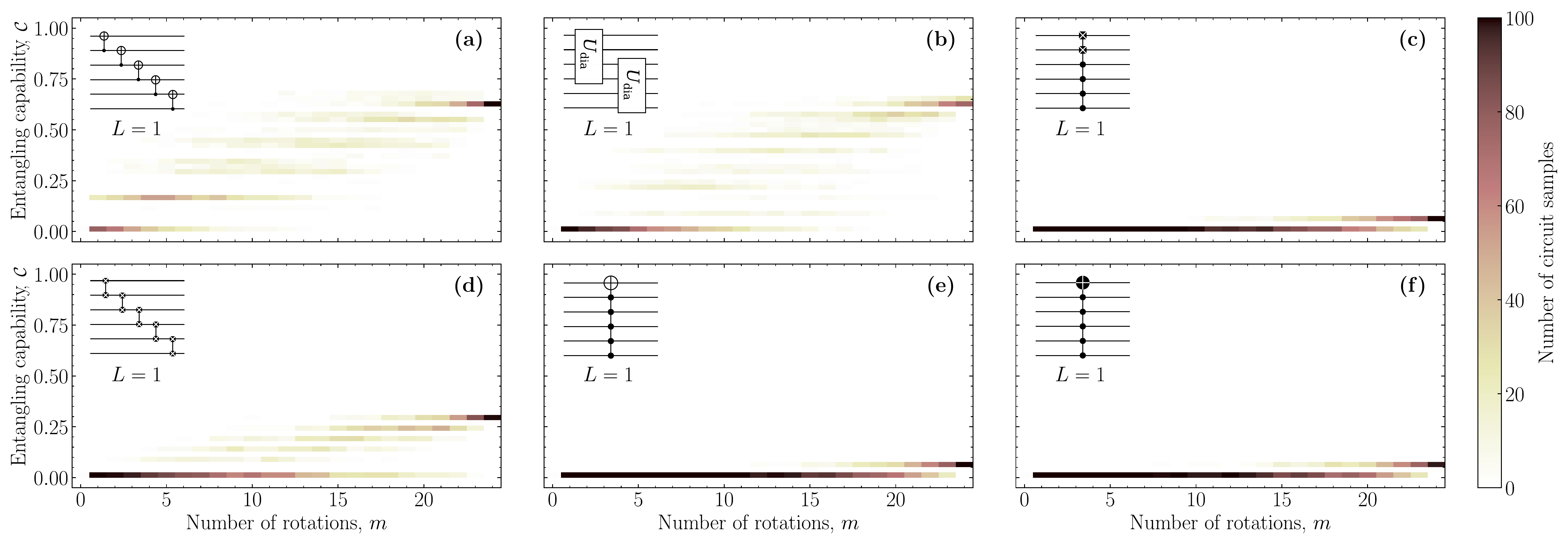}
	\caption{\emph{Entangling capability} of different circuits with $N=6$ qubits and $L=1$ layer as a function of the number of rotations. The inset in the right upper corner of each plot shows the entangling gate, $V$, used in each layer. \textbf{(a)} Five \cnot{}s, \textbf{(b)} two diamond gates, \textbf{(c)} \iswap with four control qubits, \textbf{(d)} five \iswap{}s, \textbf{(e)} \notgate with five control qubits, \textbf{(f)} $i$\notgate with five control qubits.}
	\label{fig:entL1_q6}
\end{figure}

\begin{figure}
	\includegraphics[width=\textwidth]{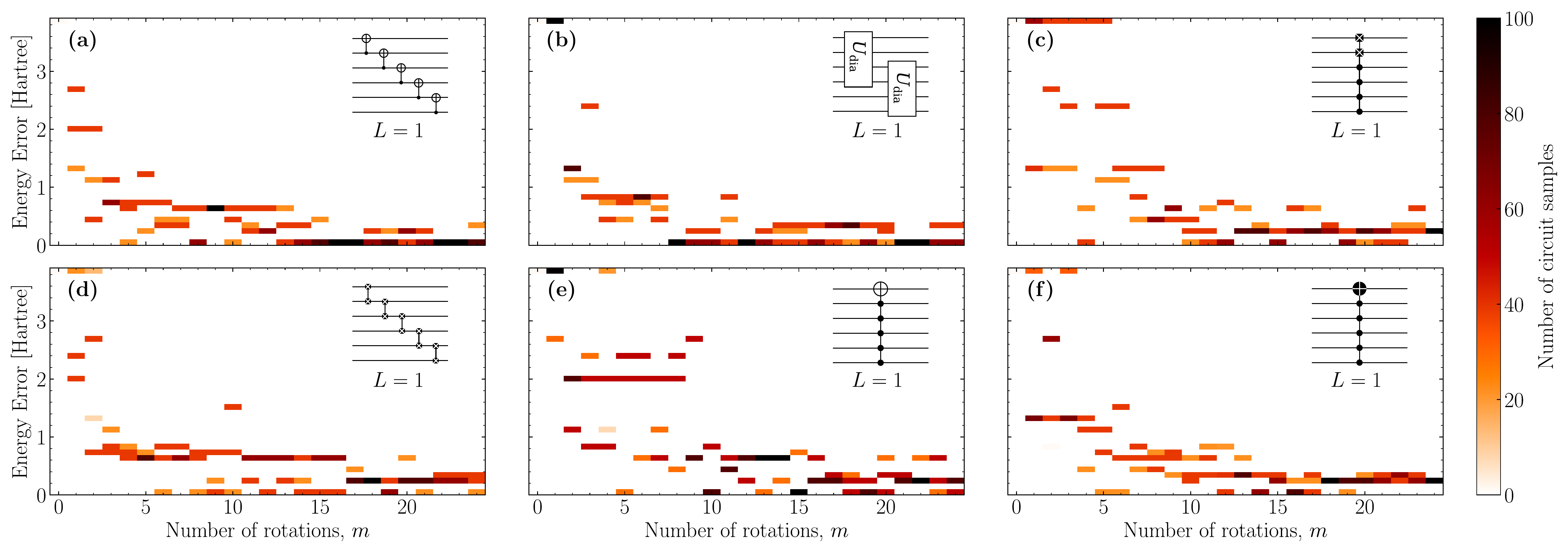}
	\caption{\emph{Energy error} of the ground state BeH$_2$ found using VQE of different circuits with $N=6$ qubits and $L=1$ layer as a function of the number of rotations. The energy found using VQE is compared to a classical diagonalization of the reduced Hamiltonian, which yields the energy error. The inset in the right upper corner of each plot shows the entangling gate, $V$, used in each layer. \textbf{(a)} Five \cnot{}s, \textbf{(b)} two diamond gates, \textbf{(c)} \iswap with four control qubits, \textbf{(d)} five \iswap{}s, \textbf{(e)} \notgate with five control qubits, \textbf{(f)} $i$\notgate with five control qubits.}
	\label{fig:vqeL1_q6}
\end{figure}

\begin{figure}
	\centering
	\includegraphics[width=\textwidth]{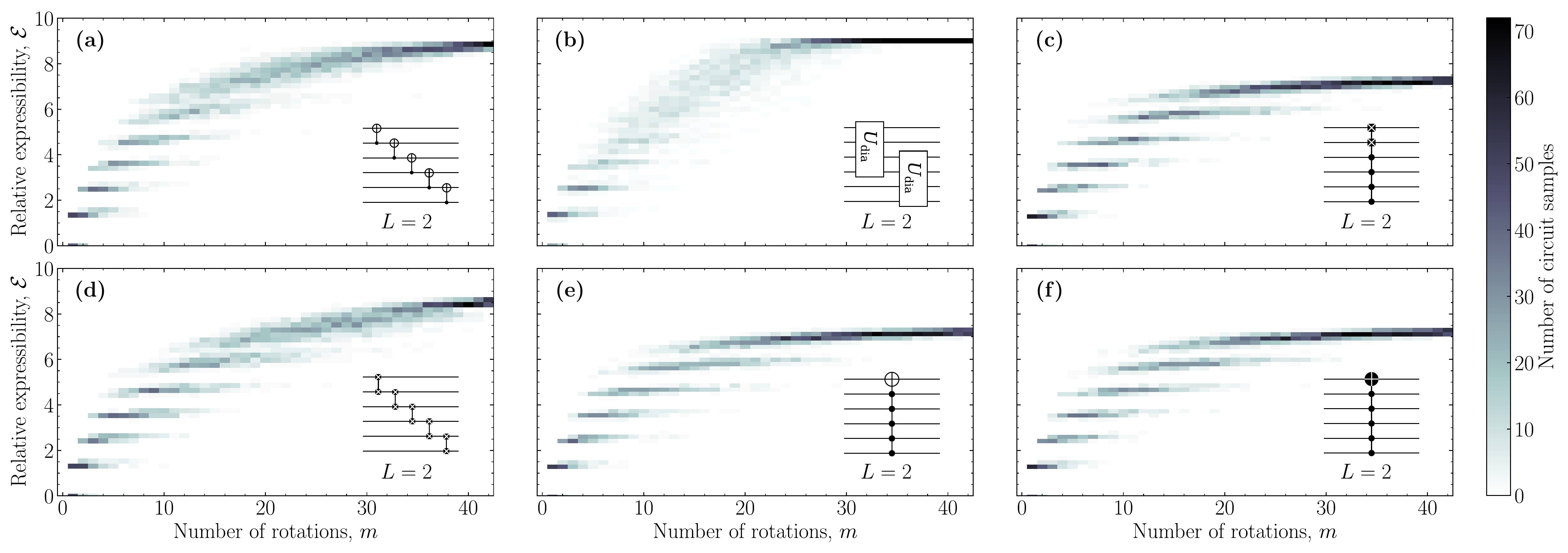}
	\caption{\emph{Relative expressibility} of different circuits with $N=6$ qubits and $L=2$ layers as a function of the number of rotations. The inset in the right upper corner of each plot shows the entangling gate, $V$, used in each layer. \textbf{(a)} Five \cnot{}s, \textbf{(b)} two diamond gates, \textbf{(c)} \iswap with four control qubits, \textbf{(d)} five \iswap{}s, \textbf{(e)} \notgate with five control qubits, \textbf{(f)} $i$\notgate with five control qubits.}
	\label{fig:exprL2_q6}
\end{figure}

\begin{figure}
	\centering
	\includegraphics[width=\textwidth]{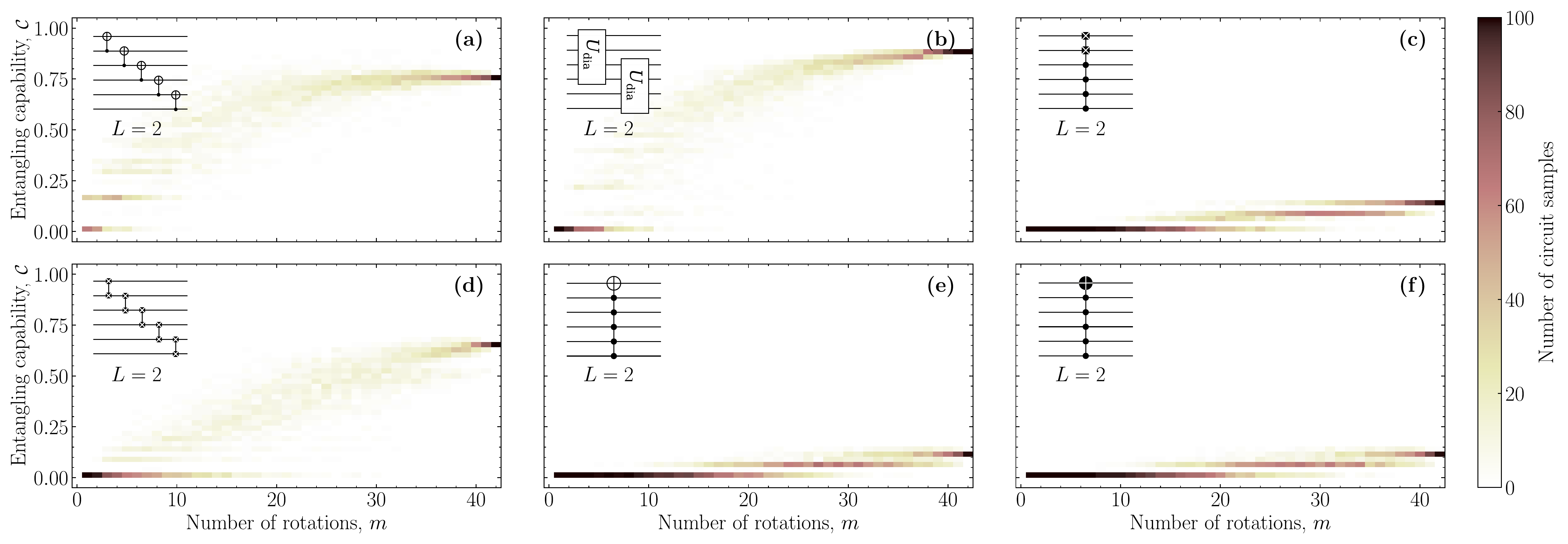}
	\caption{\emph{Entangling capability} of different circuits with $N=6$ qubits and $L=2$ layers as a function of the number of rotations. The inset in the right upper corner of each plot shows the entangling gate, $V$, used in each layer. \textbf{(a)} Five \cnot{}s, \textbf{(b)} two diamond gates, \textbf{(c)} \iswap with four control qubits, \textbf{(d)} five \iswap{}s, \textbf{(e)} \notgate with five control qubits, \textbf{(f)} $i$\notgate with five control qubits.}
	\label{fig:entL2_q6}
\end{figure}

\begin{figure}
	\includegraphics[width=\textwidth]{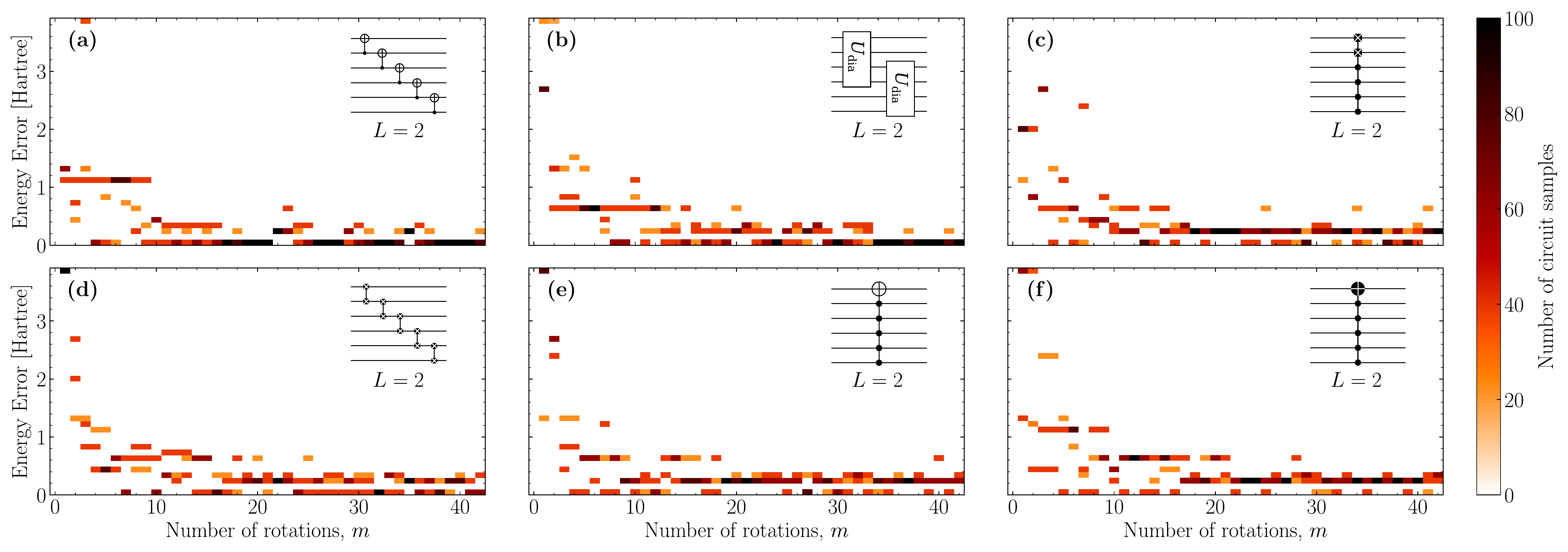}
	\caption{\emph{Energy error} of the ground state BeH$_2$ found using VQE of different circuits with $N=6$ qubits and $L=2$ layers as a function of the number of rotations. The energy found using VQE is compared to a classical diagonalization of the reduced Hamiltonian, which yields the energy error. The inset in the right upper corner of each plot shows the entangling gate, $V$, used in each layer. \textbf{(a)} Five \cnot{}s, \textbf{(b)} two diamond gates, \textbf{(c)} \iswap with four control qubits, \textbf{(d)} five \iswap{}s, \textbf{(e)} \notgate with five control qubits, \textbf{(f)} $i$\notgate with five control qubits.}
	\label{fig:vqeL2_q6}
\end{figure}

\begin{figure}
	\centering
	\includegraphics[width=\textwidth]{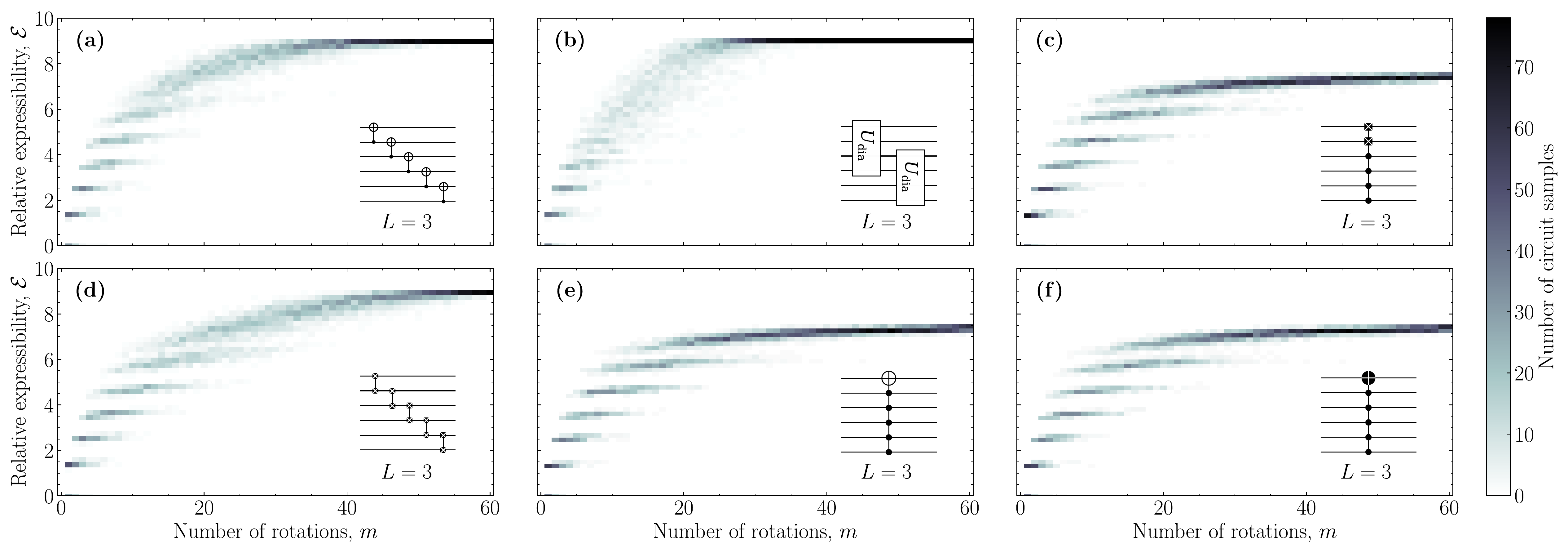}
	\caption{\emph{Relative expressibility} of different circuits with $N=6$ qubits and $L=3$ layers as a function of the number of rotations. The inset in the right upper corner of each plot shows the entangling gate, $V$, used in each layer. \textbf{(a)} Five \cnot{}s, \textbf{(b)} two diamond gates, \textbf{(c)} \iswap with four control qubits, \textbf{(d)} five \iswap{}s, \textbf{(e)} \notgate with five control qubits, \textbf{(f)} $i$\notgate with five control qubits.}
	\label{fig:exprL3_q6}
\end{figure}

\begin{figure}
	\includegraphics[width=\textwidth]{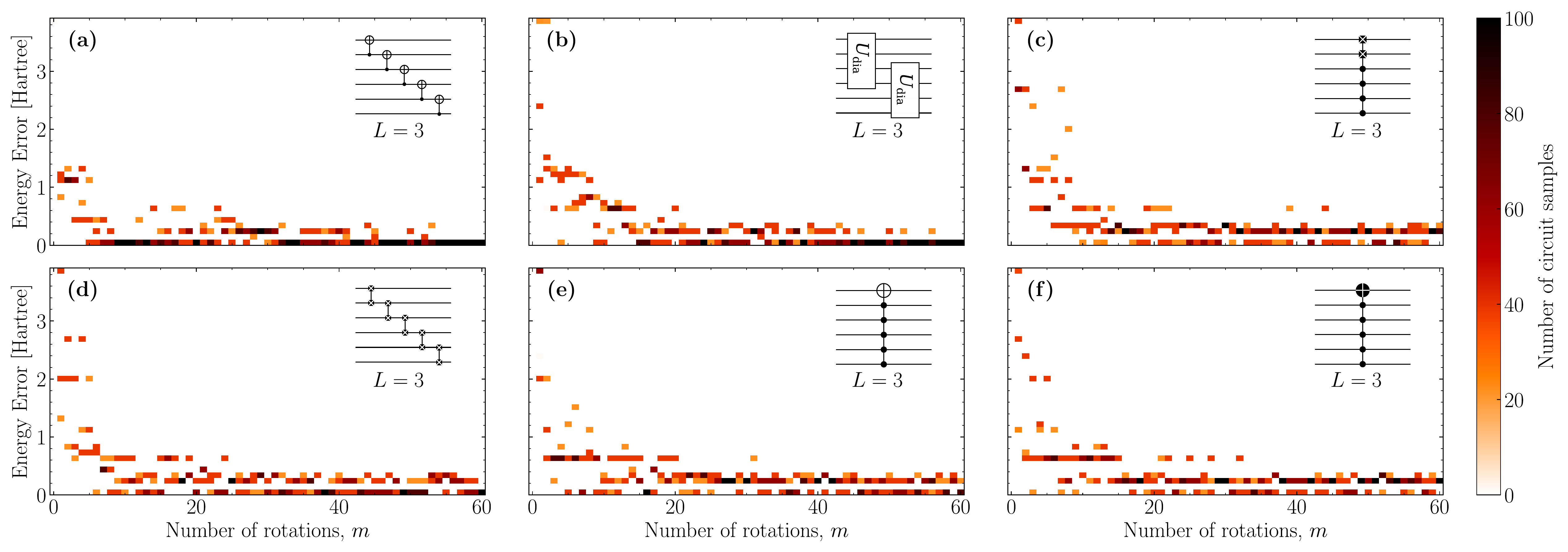}
	\caption{\emph{Energy error} of the ground state BeH$_2$ found using VQE of different circuits with $N=6$ qubits and $L=3$ layer as a function of the number of rotations. The energy found using VQE is compared to a classical diagonalization of the reduced Hamiltonian, which yields the energy error. The inset in the right upper corner of each plot shows the entangling gate, $V$, used in each layer. \textbf{(a)} Five \cnot{}s, \textbf{(b)} two diamond gates, \textbf{(c)} \iswap with four control qubits, \textbf{(d)} five \iswap{}s, \textbf{(e)} \notgate with five control qubits, \textbf{(f)} $i$\notgate with five control qubits.}
	\label{fig:vqeL3_q6}
\end{figure}

\begin{figure}
	\centering
	\includegraphics[width=\textwidth]{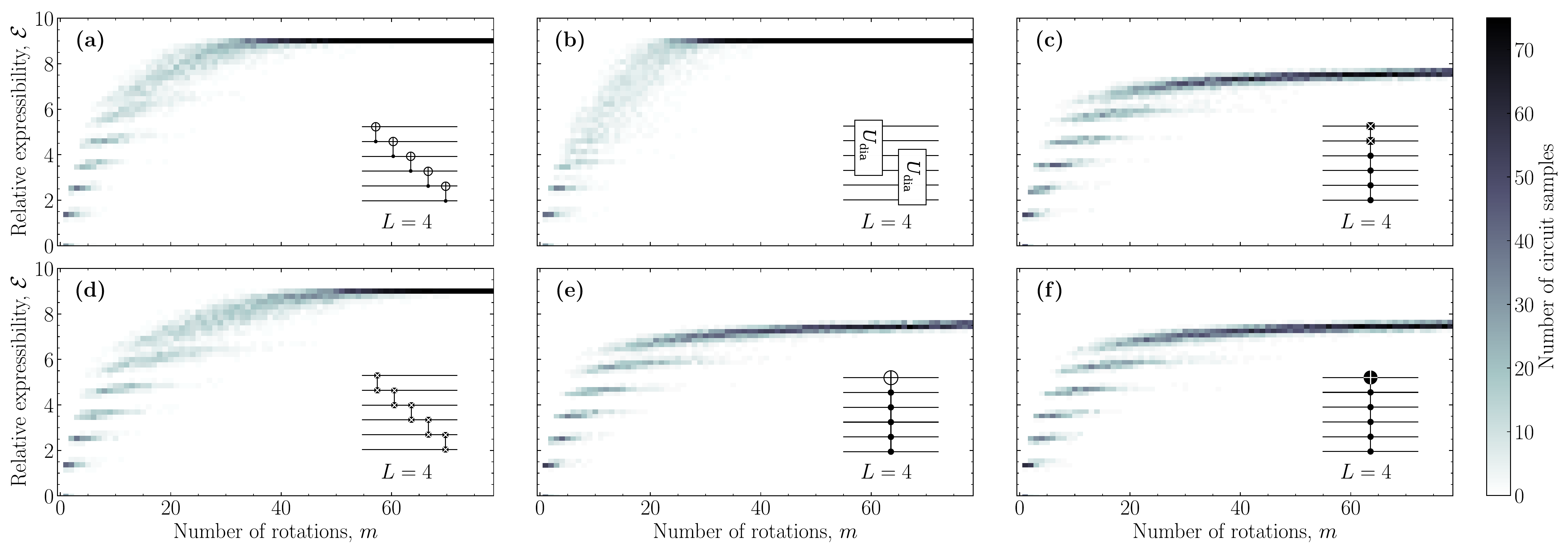}
	\caption{\emph{Relative expressibility} of different circuits with $N=6$ qubits and $L=4$ layers as a function of the number of rotations. The inset in the right upper corner of each plot shows the entangling gate, $V$, used in each layer. \textbf{(a)} Five \cnot{}s, \textbf{(b)} two diamond gates, \textbf{(c)} \iswap with four control qubits, \textbf{(d)} five \iswap{}s, \textbf{(e)} \notgate with five control qubits, \textbf{(f)} $i$\notgate with five control qubits.}
	\label{fig:exprL4_q6}
\end{figure}

\begin{figure}
	\includegraphics[width=\textwidth]{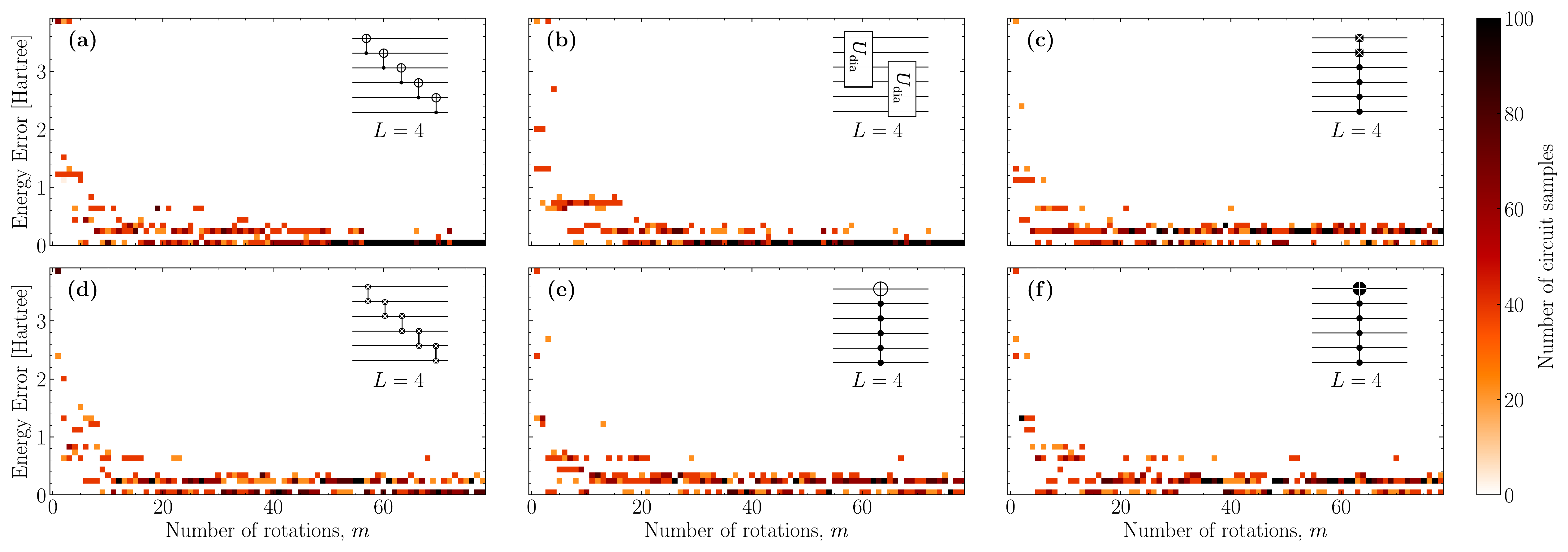}
	\caption{\emph{Energy error} of the ground state BeH$_2$ found using VQE of different circuits with $N=6$ qubits and $L=4$ layers as a function of the number of rotations. The energy found using VQE is compared to a classical diagonalization of the reduced Hamiltonian, which yields the energy error. The inset in the right upper corner of each plot shows the entangling gate, $V$, used in each layer. \textbf{(a)} Five \cnot{}s, \textbf{(b)} two diamond gates, \textbf{(c)} \iswap with four control qubits, \textbf{(d)} five \iswap{}s, \textbf{(e)} \notgate with five control qubits, \textbf{(f)} $i$\notgate with five control qubits.}
	\label{fig:vqeL4_q6}
\end{figure}

\begin{figure}
	\centering
	\includegraphics[width=\textwidth]{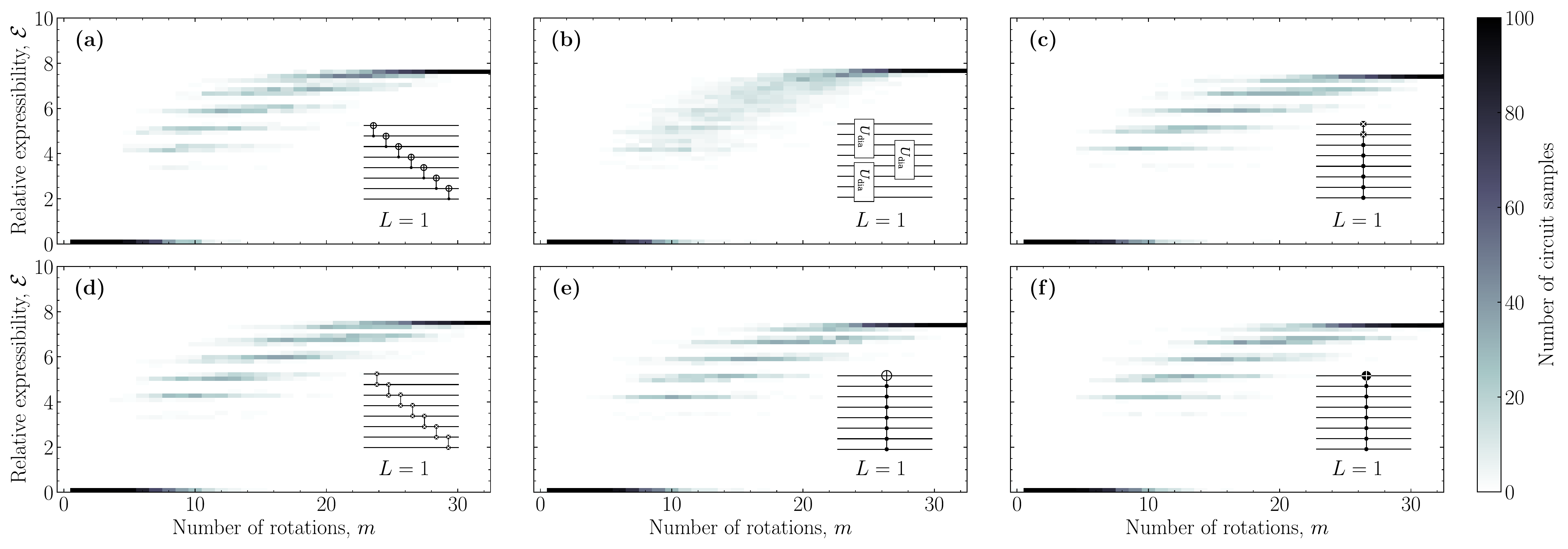}
	\caption{\emph{Relative expressibility} of different circuits with $N=8$ qubits and $L=1$ layer as a function of the number of rotations. The inset in the right upper corner of each plot shows the entangling gate, $V$, used in each layer. \textbf{(a)} Seven \cnot{}s, \textbf{(b)} three diamond gates, \textbf{(c)} \iswap with six control qubit, \textbf{(d)} seven \iswap{}s, \textbf{(e)} \notgate with seven control qubits, \textbf{(f)} $i$\notgate with seven control qubits.}
	\label{fig:exprL1_q8}
\end{figure}

\begin{figure}
	\includegraphics[width=\textwidth]{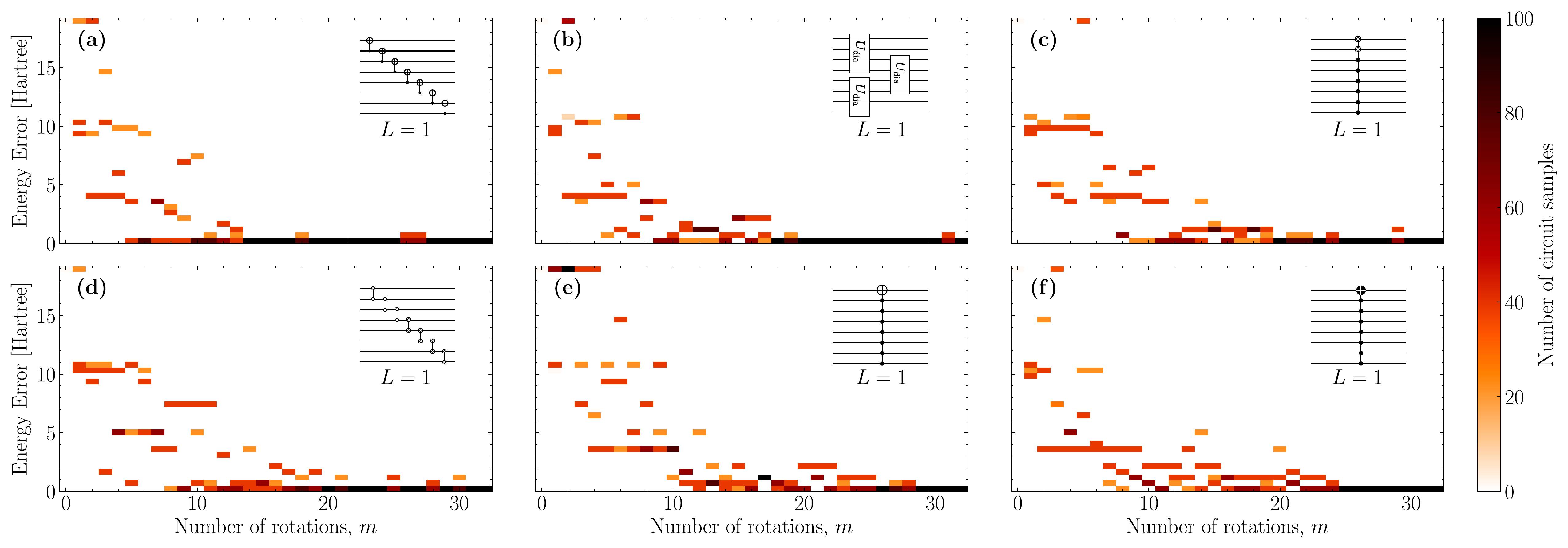}
	\caption{\emph{Energy error} of the ground state OH found using VQE of different circuits with $N=8$ qubits and $L=1$ layer as a function of the number of rotations. The energy found using VQE is compared to a classical diagonalization of the reduced Hamiltonian, which yields the energy error. The inset in the right upper corner of each plot shows the entangling gate, $V$, used in each layer. \textbf{(a)} Seven \cnot{}s, \textbf{(b)} three diamond gates, \textbf{(c)} \iswap with six control qubit, \textbf{(d)} seven \iswap{}s, \textbf{(e)} \notgate with seven control qubits, \textbf{(f)} $i$\notgate with seven control qubits.}
	\label{fig:vqeL1_q8}
\end{figure}

\begin{figure}
	\centering
	\includegraphics[width=\textwidth]{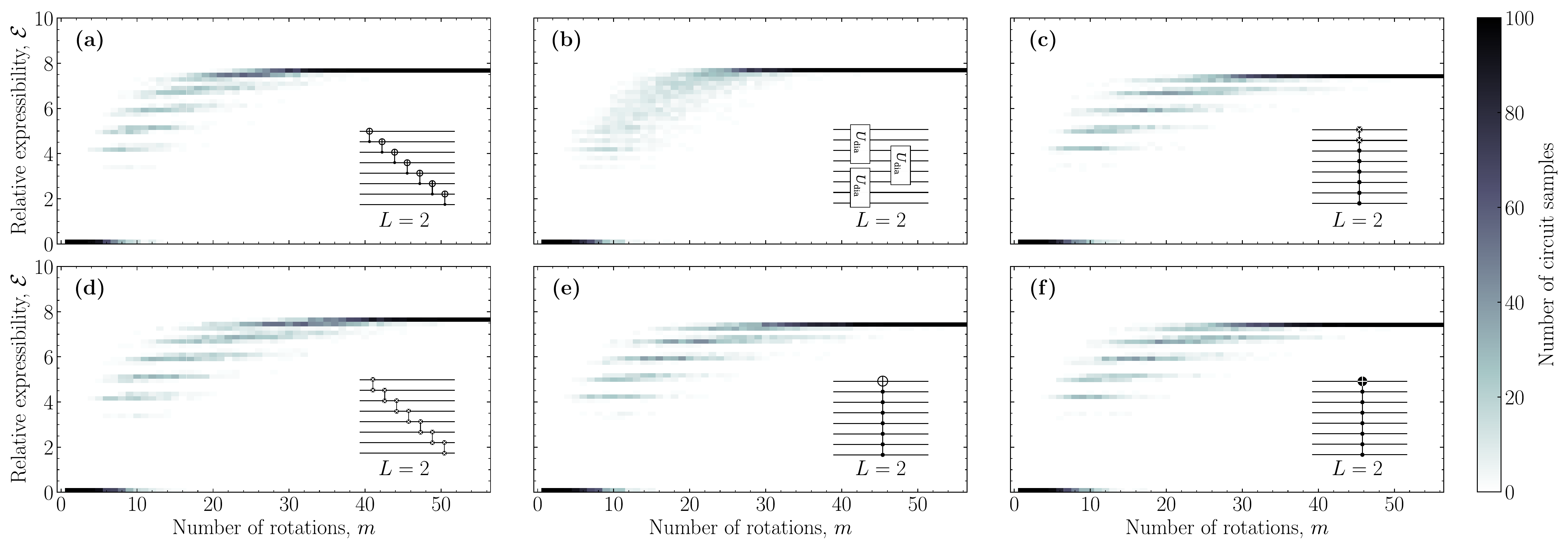}
	\caption{Relative expressibility of different circuits with $N=8$ qubits and $L=2$ layers as a function of the number of rotations. The inset in the right upper corner of each plot shows the entangling gate, $V$, used in each layer. \textbf{(a)} Seven \cnot{}s, \textbf{(b)} three diamond gates, \textbf{(c)} \iswap with six control qubit, \textbf{(d)} seven \iswap{}s, \textbf{(e)} \notgate with seven control qubits, \textbf{(f)} $i$\notgate with seven control qubits.}
	\label{fig:exprL2_q8}
\end{figure}

\begin{figure}
	\includegraphics[width=\textwidth]{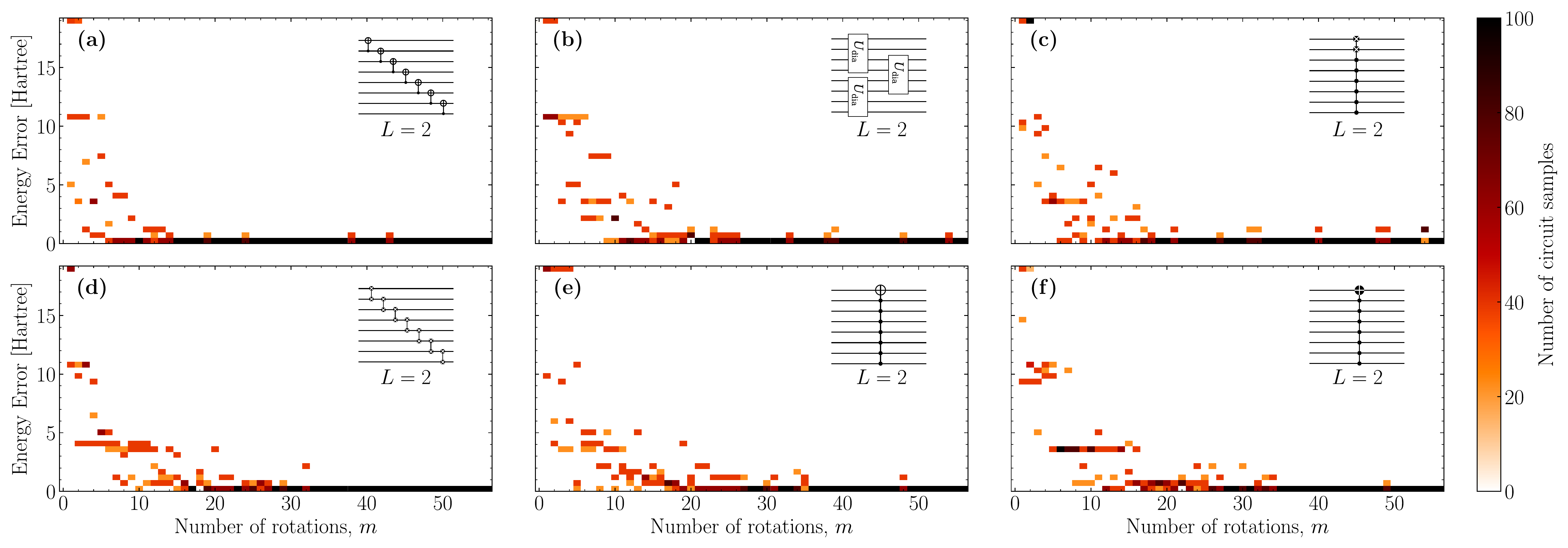}
	\caption{\emph{Energy error} of the ground state OH found using VQE of different circuits with $N=8$ qubits and $L=2$ layers as a function of the number of rotations. The energy found using VQE is compared to a classical diagonalization of the reduced Hamiltonian, which yields the energy error. The inset in the right upper corner of each plot shows the entangling gate, $V$, used in each layer. \textbf{(a)} Seven \cnot{}s, \textbf{(b)} three diamond gates, \textbf{(c)} \iswap with six control qubit, \textbf{(d)} seven \iswap{}s, \textbf{(e)} \notgate with seven control qubits, \textbf{(f)} $i$\notgate with seven control qubits.}
	\label{fig:vqeL2_q8}
\end{figure}

\begin{figure}
	\centering
	\includegraphics[width=\textwidth]{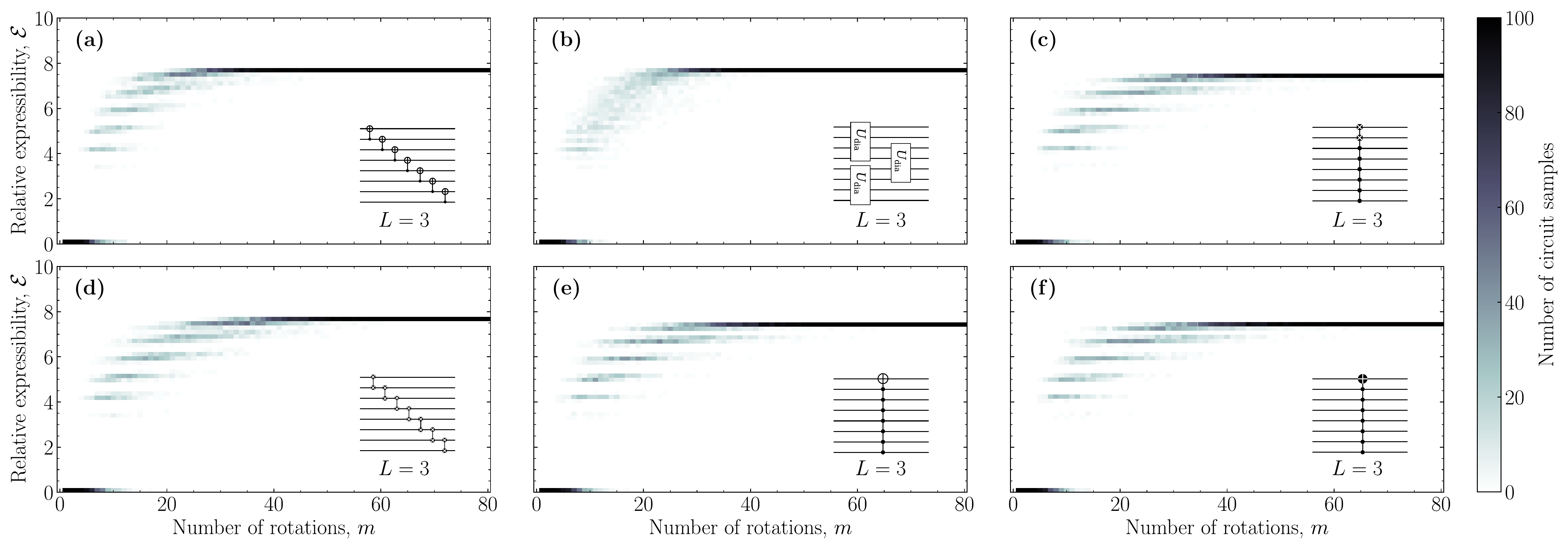}
	\caption{Relative expressibility of different circuits with $N=8$ qubits and $L=3$ layers as a function of the number of rotations. The inset in the right upper corner of each plot shows the entangling gate, $V$, used in each layer. \textbf{(a)} Seven \cnot{}s, \textbf{(b)} three diamond gates, \textbf{(c)} \iswap with six control qubit, \textbf{(d)} seven \iswap{}s, \textbf{(e)} \notgate with seven control qubits, \textbf{(f)} $i$\notgate with seven control qubits.}
	\label{fig:exprL3_q8}
\end{figure}

\begin{figure}
	\includegraphics[width=\textwidth]{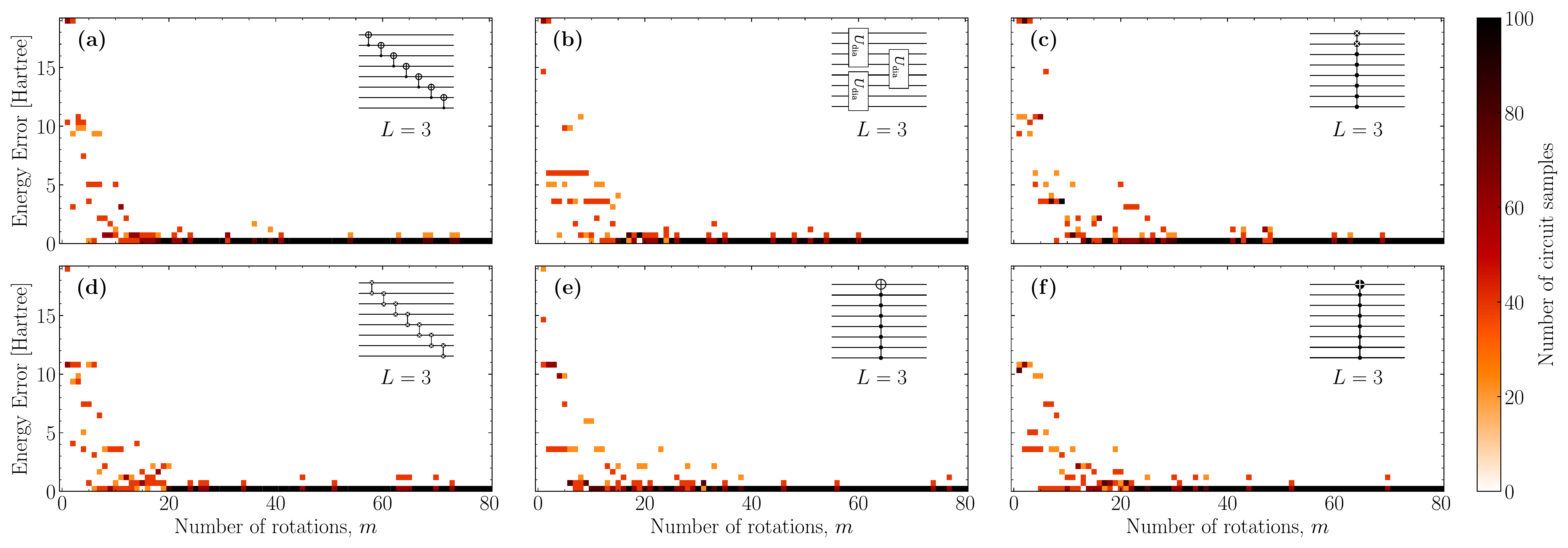}
	\caption{\emph{Energy error} of the ground state OH found using VQE of different circuits with $N=8$ qubits and $L=3$ layers as a function of the number of rotations. The energy found using VQE is compared to a classical diagonalization of the reduced Hamiltonian, which yields the energy error. The inset in the right upper corner of each plot shows the entangling gate, $V$, used in each layer. \textbf{(a)} Seven \cnot{}s, \textbf{(b)} three diamond gates, \textbf{(c)} \iswap with six control qubit, \textbf{(d)} seven \iswap{}s, \textbf{(e)} \notgate with seven control qubits, \textbf{(f)} $i$\notgate with seven control qubits.}
	\label{fig:vqeL3_q8}
\end{figure}

\clearpage